%% file: main_arxiv.tex
\documentclass[fleqn,usenatbib]{mnras}
\usepackage{newtxtext,newtxmath, import}
\usepackage{graphicx, amsmath, bbm, mathtools, capt-of, orcidlink} 
\usepackage[para]{threeparttablex}
\usepackage{eqnarray}

\usepackage[T1]{fontenc}
\usepackage[dvipsnames]{xcolor}
\usepackage[switch]{lineno}

\title[Flux-ratio prediction for quadruply lensed quasars with curved arc models]{Macromodel-free flux-ratio prediction in quadruply imaged quasars with local constraints from lensed arcs}

\author[H. Paugnat et al.]{
Hadrien Paugnat \orcidlink{0000-0002-2603-6031},$^{1}$\thanks{E-mail: hpaugnat@astro.ucla.edu}
Tommaso Treu \orcidlink{0000-0002-8460-0390},$^{1}$
and
Daniel Gilman \orcidlink{0000-0002-5116-7287}$^{2}$\thanks{Brinson Prize Fellow}
\\
$^{1}$Department of Physics and Astronomy, UCLA, Los Angeles, CA 90095-1547, USA\\
$^{2}$Department of Astronomy \& Astrophysics, University of Chicago, Chicago, IL 60637, USA\\
}

\date{\vspace{-1.5cm}}
\pubyear{\the\year{}}

\begin{document}
\label{firstpage}
\pagerange{\pageref{firstpage}--\pageref{lastpage}}
\maketitle

\begin{abstract}
\import{}{abstract.tex}
\end{abstract}

\begin{keywords}
gravitational lensing: strong--cosmology: dark matter--galaxies: structure
\vspace{-0.05cm}
\end{keywords}

\import{}{body_text.tex}

\vspace{-0.45cm}
\section*{Acknowledgements}
\vspace{-0.15cm}

We thank Simon Birrer, Anna Nierenberg, and Xiaolong Du for helpful discussions, comments and feedback. This work was supported by the National Science Foundation under grant AST-2205100, and by the Gordon and Betty Moore Foundation under grant No. 8548. DG acknowledges support provided by the Brinson Foundation through a Brinson Prize Fellowship grant. 

This research made use of the following software packages: \texttt{lenstronomy} \citep{Birrer2015,lenstronomy, lenstronomy2}, \texttt{pyHalo}  \citep{Gilman2020}, \texttt{paltas} \citep{Wagner-Carena2023}, and \texttt{corner} \citep{corner}.
Some of the calculations were performed using the Hoffman2 Shared Cluster, provided by the UCLA Institute for Digital Research and Education’s Research Technology Group.

\vspace{-0.45cm}
\section*{Data Availability}
\vspace{-0.15cm}

The simulated data used in this article were created with \texttt{lenstronomy} commit 469e38f, which is openly available in the main public code repository at \url{https://github.com/lenstronomy/lenstronomy}. Simulated imaging data discussed in the manuscript are not publicly available upon publication because is not technically feasible and/or the cost of preparing, depositing, and hosting the data would be prohibitive within the terms of this research project. The data are available from the authors upon reasonable request.

\vspace{-0.45cm}
\bibliographystyle{mnras}
\bibliography{biblio} 


\appendix
\import{}{appendices.tex}


\bsp	
\label{lastpage}
\end{document}

%% file: abstract.tex
Strong gravitational lensing is a powerful cosmological probe, providing a direct tool to unveil the properties of dark matter (DM) on subgalactic scales. In particular, flux-ratio anomalies in quadruply imaged quasars (“quads”) can reveal the presence of dark substructure, such that population-level statistics can be used to constrain the particle nature of DM. Current methods, however, rely on globally parametrized models (“macromodels”) of the lens mass distribution, which impose rigid physical assumptions on the deflection field. Given the high stakes, it is important to develop complementary methods that do not require the assumption of a macromodel. One promising avenue consists of modeling the resolved emission from the quasar host galaxy (lensed arcs) using a local lensing formalism like the Curved Arc Basis (CAB) description. This method effectively imposes minimal priors on the mass distribution within the lens galaxy and thus will provide conservative error estimates on the inferred DM properties.
In this paper, we test the ability of CAB models to predict flux ratios from mock imaging data. We find that the CAB model-predicted flux ratios accurately reproduce the expected values, with a typical precision of $\sim\text{3-5}\%$. While a macromodel-based approach yields smaller uncertainties, as expected, the CAB method permits a more flexible, local description of the deflection field, thus being more robust to angular structure in the main deflector mass profile, in particular avoiding false-positive detections of flux-ratio anomalies that can arise with overly simplistic parametrizations.
On the other hand, by injecting individual DM halos near quasar images, we demonstrate that CAB models do not absorb the local lensing perturbations from DM substructure, and can therefore distinguish flux-ratio anomalies caused by DM substructure from other sources of small-scale perturbation. We conclude that CAB lens models can be used to infer DM properties from flux-ratio anomaly statistics with minimal assumptions, complementing the traditional macromodel-based approach.

%% file: body_text.tex
\section{Introduction} \label{sec:intro}

The standard model of cosmology posits that $\sim 5/6$ of the universe's matter content consists of cold dark matter (CDM), a collisionless particle that only interacts weakly with ordinary baryonic matter and radiation, predominantly through gravity. CDM successfully explains a wide range of astrophysical observations, in particular the cosmic microwave background \citep[e.g.,][]{Planck2020}, the formation of cosmic structures on scales $\gtrsim 0.1$ Mpc \citep[e.g.,][]{Tegmark2004} (mass scales $\gtrsim 10^{11} M_\odot$), and the profiles of dark matter halos large enough to host observable galaxies \citep[e.g.,][]{Weinberg2015}. The CDM paradigm, however, faces two challenges that have driven the development of alternative theories: first, the lack of detection of a viable CDM particle candidate despite decades of experimental efforts \citep[e.g.,][]{Feng2010, Schumann2019}; second, a potential disagreement between predictions and observations at smaller, subgalactic scales \citep[e.g.,][]{Bullock2017} - though a deeper comprehension of the baryonic processes at play might alleviate this tension.

To better understand the nature of dark matter (DM), and to discriminate between CDM and alternative paradigms, it is therefore necessary to characterize the properties of substructure on “small” ($\lesssim 10^{10} M_\odot$) scales. Several observational probes can be exploited to this effect: dwarf galaxy surveys, and in particular of the Milky Way satellite galaxy population \citep[e.g.,][]{Nadler2021, Dekker2022, Nadler2024}, measurements of the Lyman-alpha forest power spectrum \citep[e.g.,][]{Viel2013, Irsic2017, Irsic2024, Villasenor2023}, dynamical effects on stellar streams in the Local Group \citep[e.g,][]{Erkal2015, Banik2021, Carlberg2024}, and gravitational lensing, both weak \citep[e.g.,][]{Mondino2024} and strong \citep[for a recent review, see][]{Vegetti2024}. More stringent constraints can be obtained by combining these different probes to break degeneracies \citep[e.g,][]{Nadler2021_joint}.

Gravitational lensing is a particularly powerful method, since it does not require accurate modeling of baryonic physics: being sensitive to the entire line-of-sight mass distribution, this effect can directly measure the gravitational signal of DM substructure. In strongly lensed systems, a massive foreground object (typically, a galaxy or galaxy cluster) bends the light from a more distant source, resulting in multiple images. Quadruply imaged quasars, with their four distinct light paths due to a galaxy-scale lens, are especially interesting targets, since the positions of the quasar images provide precise constraints for the smooth mass distribution in the lens. By detecting localized gravitational perturbations of this smooth mass distribution, one can infer the presence of substructure, either subhalos within the DM halo of the main deflector \citep[e.g.,][]{Mao1998}, or field halos along the line-of-sight which are close (in projection) to the lensed images \citep[e.g.][]{Despali2018}. 


The most straightforward approach - direct detection - is to look for observable perturbations of individual DM halos in a given lens system. This approach has led to a small number of significant detections \citep[e.g.][]{Vegetti2010, Vegetti2012, Nierenberg2014, Hezaveh2016, Nightingale2024}, allowing to probe the inner structure of subhalos \citep[e.g.,][]{Minor2021a, Minor2021b, Despali2025}. However, this method is only sensitive to the most massive subhalos \citep[$\gtrsim10^9 M_\odot$, e.g.][]{Ritondale2019}, requires involved computations, and spurious low-significance detections or systematic errors in the recovered subhalo properties can be caused by  assumptions in the source light reconstruction \citep[e.g.][]{Ballard2024, Ephremidze2025} or the lens mass model \citep[e.g.][]{Nightingale2024, O'Riordan2024, Lange2025, Stacey2025}.

Rather than attempting to detect individual halos, an alternative approach - statistical detection - consists of connecting the subgalactic properties of DM to the collective imprint left by dark substructure at the population level. Direct detections and non-detections in a sample of observed lenses, for instance, can be used to constrain the subhalo mass function \citep[e.g.,][]{Ritondale2019}, but this requires careful modeling of the (quite complex) sensitivity \citep[e.g.,][]{Despali2022, O'Riordan2023}. Machine learning techniques have been developed to estimate substructure properties in a more agnostic way, directly from the images \citep[e.g.][]{Ostdiek2022,Wagner-Carena2023, Zhang2024}, a framework that can be scaled to large datasets, but with the limitation that employing simplified simulations for the training sets is likely to produce a biased inference.

Statistical detection can otherwise be carried out by reducing the images to meaningful low-dimensional summary statistics. In particular, “flux-ratio anomalies” \citep[discrepancies between the observed flux ratios between the quasar images and the relative magnifications expected from the smooth mass distribution, ][]{Mao1998, MetcalfMadau2001, Dalal2002} can reveal the presence of substructure down to $\sim 10^6-10^7 M_\odot$ \citep[e.g.,][]{Nierenberg2017, Gilman2019, Nierenberg2024}. Population-level statistics of these flux-ratio anomalies have in consequence been exploited to place constraints on potential alternatives to the CDM paradigm : a non-exhaustive list includes Warm Dark Matter \citep[WDM; e.g.,][]{Birrer2017, Gilman2020,Hsueh2020, Keeley2024}, sterile neutrinos \citep{Zelko2022}, primordial black holes \citep[e.g.,][]{Dike2023}, and Self-Interacting Dark Matter \citep[SIDM ; e.g.,][]{Gilman2021,Gilman2023}.


To predict the relative magnifications that are used to detect flux-ratio anomalies, the baseline approach consists of assuming a globally parametrized profile (often called a “macromodel”) for the deflection field of the main deflector, motivated by physical expectations about the smooth mass distribution, and constraining the model parameters with the precisely measured image positions of the quasar \citep[e.g.,][]{Mao1998, Dalal2002, Nierenberg2014, Nierenberg2017, Gilman2020, Keeley2024}. Thanks to the substantial improvement of galaxy-scale lensing observations over the past decades, extended lensed arcs due to the quasar host galaxy can now be observed and leveraged to better constrain the lens mass distribution \citep[e.g.,][]{Oh2024}. Recent technical developments enable to simultaneously reconstruct lensed arcs and flux ratios in the presence of an entire substructure population, such that DM constraints with flux-ratio anomaly statistics can be strengthened by incorporating the full imaging data \citep{Gilman2024}.

One may be concerned, however, that the macromodel approach may be oversimplistic, since it imposes physical assumptions about the global shape of the deflection field. Similarly to the direct detection case, unaccounted-for complexity - such as deviations from ellipticity \citep[e.g.,][]{Congdon2005, Powell2022, Cohen2024}, or an edge-on baryonic disk component \citep{Hsueh2016, Hsueh2017, Gilman2017} - or even improper parametrizations of the lens mass distribution \citep[e.g.,][]{Paugnat2025} can produce false-positive flux-ratio anomalies in individual systems. At a statistical level, there is a risk that this might underestimate the uncertainties and/or introduce systematic errors in the distribution of flux-ratio anomalies \citep[e.g.,][]{Gilman2017, Hsueh2018, Gilman2024, Paugnat2025}. Recent studies \citep[e.g.,][]{Gilman2022,Keeley2024} mitigate this concern by adding extra degrees of freedom to the baseline mass models (e.g., multipoles, see Section~\ref{subsec:macromodels}), and avoiding complex deflectors as evaluated from high-resolution imaging. However, there is no guarantee that the lens properties are fully captured with any imposed functional form. Therefore, in order to obtain conclusive evidence about the nature of DM, it is important to develop methods that do not rely on the assumption of a macromodel.

A promising idea to bypass the need for a macromodel is to characterize the lensing properties \textit{locally}, instead of assuming a structure for the deflection field over the entire image.
\cite{Birrer_CAB} proposed the Curved Arc Basis (CAB) formalism as a flexible, macromodel-free description of the gravitational distortions in the vicinity of an extended lensed arc, based on the eigenvalues and eigenvectors of the local lensing Jacobian and their directional differentials. While some recent work \citep{Sengul2023, Ephremidze2025} develops this novel approach in the context of cluster lenses, it has yet to be applied to DM inference with galaxy-scale lenses.

This paper is intended to be a proof-of-concept for CAB models in the context of flux-ratio anomaly statistics of quadruply imaged quasars with extended lensed arcs. We present a practical implementation of the idea, and test its capacity to accurately predict the dark substructure-free flux ratios on simulated data. We compare its performance to a macromodel-based approach, using various simulation setups to verify that the enhanced flexibility of CAB models produces flux-ratio predictions that are robust to complexity in the lens mass distribution, while still enabling the detection of flux-ratio anomalies from substructure.
This is intended to be the first building block of an effort to use CAB models for a full macromodel-free inference of DM properties with flux-ratio anomaly statistics.

Taking a broader view, the macromodel and CAB formalisms can be thought of as bookends to the “true correct model”: whereas the macromodel approach may be underestimating the uncertainties by imposing too little flexibility in the modeling of the deflector mass distribution, the CAB approach may be overestimating them by allowing too much flexibility. After all, the majority of galaxy-scale deflectors are massive elliptical galaxies, which are known to be very regular and self-similar \citep[e.g.][]{Auger2010,Cappellari2016}. By applying and comparing the results from both formalisms we can strengthen our confidence that the inference of DM properties is robust and that the error estimates are realistic.

This paper is organized as follows. In Section~\ref{sec:formalism}, we provide a brief review of the macromodel and CAB approaches to lens modeling. In Section~\ref{sec:test}, we present a framework to comparatively test the ability of these two methods to predict flux ratios in simulated datasets of quadruply imaged quasars with extended lensed arcs. In Section~\ref{sec:results}, we present the results of these tests, showing that CAB models can detect actual flux-ratio anomalies but are less susceptible to false positives than a given macromodel. We summarize our findings in Section~\ref{sec:discussion} and offer perspective for future applications.

\section{Curved arc basis formalism} \label{sec:formalism}

In this section, we first give a reminder of the general gravitational lensing formalism (Section~\ref{subsec:grav_lensing_formalism}), before discussing the usual parametrized macromodel approach to lens modeling, and describing some frequently used profiles(Section~\ref{subsec:macromodels}). Finally, we review the CAB formalism, introduced by \cite{Birrer_CAB} to describe extended arcs by extrapolating local lensing properties rather than assuming a global parametrization (Section~\ref{subsec:CAB_method}).

\subsection{Gravitational lensing formalism}
\label{subsec:grav_lensing_formalism}

In the usual gravitational lensing formalism, the true (unlensed) angular position $\vec{\beta}$ of a source and its observed angular position $\vec{\theta}$ in the image plane are related by the simple lens equation
\begin{equation} \label{eq:lens_equation}
  \vec{\beta} = \vec{\theta} - \vec{\alpha}(\vec{\theta}),
\end{equation}
where, in the case of a single thin lensing plane, the reduced deflection angle $\vec{\alpha}$ can be directly expressed as a function of the convergence (dimensionless surface mass density) profile $\kappa$ with:
\begin{equation}\label{eq:alpha_kappa_relation}
\vec{\alpha}(\vec{\theta }) =  \frac {1}{\pi } \int d^{2}\theta ^{\prime } \frac{(\vec{\theta }-\vec{\theta }^{\prime }) \kappa (\vec{\theta }^{\prime })}{ \big|\vec{\theta }-\vec{\theta }^{\prime }\big|^2 }.
\end{equation}
In that case, we can also define a scalar lensing potential 
\begin{equation}\label{eq:psi_kappa_relation}
\psi(\vec{\theta }) =  \frac {1}{\pi } \int d^{2}\theta ^{\prime } \kappa (\vec{\theta }^{\prime }) \ln \big|\vec{\theta }-\vec{\theta }^{\prime }\big|
\end{equation}
such that $\vec{\alpha}(\vec{\theta }) = \vec{\nabla}\psi(\vec{\theta })$ and $\Delta\psi(\vec{\theta }) = 2\kappa(\vec{\theta })$.

The coordinate mapping between the lensed image plane position $\vec{\theta}=(\theta_x,\theta_y)$ and the unlensed source plane position $\vec{\beta}=(\beta_x,\beta_y)$ is associated with a Jacobian matrix $\mathbf{A}$ defined by:
\begin{equation} 
A_{ij} \equiv \frac{\partial \beta_i}{\partial \theta_j}=\delta_{ij} - \frac{\partial \alpha_i}{\partial \theta_j} =\delta_{ij} - \frac{\partial^2 \psi}{\partial \theta_i\partial \theta_j}.
\label{eq:Jacobian_def}
\end{equation}
The last equality implies that the Jacobian is symmetric ($A_{ij}=A_{ji}$), so it can be decomposed in a diagonal component involving the convergence $\kappa=1-\frac{1}{2}(A_{xx}+A_{yy})$ and a trace-free term with the shear components $\gamma_1=\frac{1}{2}(A_{yy}-A_{xx})$, $\gamma_2=-A_{xy}$:
\begin{equation}
\mathbf{A} = \left[\begin{array}{ c c } 1-\kappa -\gamma_1 & -\gamma_2 \\ -\gamma_2 & 1-\kappa +\gamma_1 \end{array}\right].
\end{equation}
Since gravitational lensing preserves surface brightness \citep[e.g.,][]{Bartelmann2010}, the magnification of a source is given by the change in differential area between the source plane and the image plane, which can be expressed as:
\begin{equation}
    \mu = \frac{1}{\det(\mathbf{A})} = \frac{1}{(1-\kappa)^2-\gamma_1^2-\gamma_2^2}.
\end{equation}

\subsection{Parametrized macromodels}
\label{subsec:macromodels}
One of the most widespread approaches for galaxy-scale lens modeling is to assume globally parametrized profiles for the convergence and shear fields, with the convergence representing the large-scale smooth surface mass density distribution of the lensing galaxy. The parameters of this macromodel are then fit to best reproduce the observed quantities (image positions, or full imaging data).

Massive early-type galaxies represent the majority of known lenses (due to their larger lensing cross-section), and they are well modeled by elliptical mass/convergence profiles \citep[e.g.,][]{Turner1984,Chae2003, Yoo2005, Schneider2006, Auger2009}. A popular choice of parametrization is the Elliptical Power Law (EPL) profile \citep{TessoreMetcalf}, also known as the Power-law Elliptical Mass Distribution (PEMD):
\begin{equation}
    \label{eq:EPL}
    \kappa_{\rm EPL}(\vec{\theta}) = \frac{2-t}{2} \left( \frac{\theta_E}{\sqrt{q\theta_1^2+\theta_2^2/q}} \right)^{t},
\end{equation}
where $0<t<2$ is the logarithmic slope of the projected convergence profile, $\theta_E$ is the Einstein radius, $0<q\leq1$ is the axis ratio, and the coordinates $(\theta_1, \theta_2)$ are rotated with respect to the reference coordinates $(\theta_x, \theta_y)$ by an orientation angle $\phi_{\rm EPL}$ such that $\theta_1$ is along the semi-major axis. The circular version of this convergence profile arises from a spherically symmetric three-dimensional (3D) mass distribution with $\rho(r_{\rm 3D}) \propto r_{\rm 3D}^{-t-1}$. In particular, $t=1$ corresponds to a Singular Isothermal Ellipsoid (SIE) profile \citep{Kassiola1993,Kormann1994}, also frequently employed since lensing galaxies are typically near-isothermal \citep{Auger2010,Shajib2021}. Lens models commonly introduce an additional external (constant) shear component \citep[e.g.,][]{Kormann1994_shear,Keeton1997, WittMao1997}, meant to represent deformations of the deflection field introduced by sources other than the main lens, for instance nearby massive galaxies or large-scale structure. These pure shear distortions are described by a lensing potential

\begin{equation}
    \psi_{\rm ext} (\vec{\theta}) = \frac{\gamma_{\rm ext}}{2} \left[ (\theta_x^2-\theta_y^2) \cos ( 2\phi_{\rm ext}) + 2\theta_x\theta_y \sin( 2\phi_{\rm ext}) \right]
\end{equation}
where $\gamma_{\rm ext}, \phi_{\rm ext}$ are the absolute shear strength and orientation (in polar coordinates) with respect to the $x$-axis, respectively.

In recent years, however, with the improvement in data quality for galaxy-scale lensing observations, some shortcomings of the widespread EPL+shear macromodel have started to appear, as this simple parametrization fails to capture the complexity of the lens mass distribution. For example, the values for the shear found in strong lens models do not match theoretical expectations \citep{Keeton1997, Hilbert2007}, direct modeling of line-of-sight perturbers \citep{Moustakas2007, Wong2011} or independent weak lensing measurements \citep{Etherington2024, Hogg2025}, suggesting that these parameters take non-physical values to compensate for the lack of flexibility of the EPL profile. There is a concern that this might bias cosmological measurements relying on strong lens modeling, whether in the context of time-delay cosmography \citep{VdV2022, VdV2022_twisting}, flux-ratio anomaly statistics \citep{Cohen2024, Gilman2024, Keeley2024}, or direct detection of substructure in individual systems, where false positives can occur if the mass model is overly simplistic \citep[][]{He2023, O'Riordan2024, Lange2025, Stacey2025}. 

Recent studies attempt to mitigate this insufficient complexity in the lens models by adding deviations from ellipticity in the form of Fourier-like angular perturbations called “multipoles” \citep{Chu2013, Xu2015, Oh2024, Paugnat2025}. This addition is motivated by studies of the optical/infrared surface photometry of massive elliptical galaxies \citep[e.g.,][]{Carter1978, Bender1988, Hao2006, Chaware2014}, where isophotal shapes are found to deviate from ellipses in a similar way - with percent-level multipole amplitudes, and a dominant effect of the $m=4$ order which encodes boxyness/diskyness.  
\cite{Paugnat2025} showed that the common “circular multipole” formulation employed in strong lensing applications, expressed in polar coordinates, is only suited for near-circular systems. For systems with non-negligible ellipticities, it produces unphysical perturbation patterns that can lead to biased values in the model-predicted flux ratios when the multipole parameters are constrained from imaging data, and that, otherwise, typically overestimate the amplitude of flux-ratio perturbations from multipoles at a statistical level. Therefore, we employ the “elliptical multipoles” proposed in \cite{Paugnat2025}, which generalize the perturbations to any axis ratio $q$:

\begin{equation}
\begin{split}
    \kappa_m(\vec{\theta};q) &= \frac{\theta_E}{2R}  a_m\cos \left[m(\varphi-\varphi_m)\right], \\
    \text{where } R &= \sqrt{q\theta_1^2 +\theta_2^2/q} \ \text{ and } \varphi = \arctan(q\theta_1,\theta_2)
\end{split}
\label{eq:ell_multipole}
\end{equation}
are coordinates that depend explicitly on the axis ratio (in particular, $\varphi$ is the eccentric anomaly, and not the polar angle).
Eq.~(\ref{eq:ell_multipole}) is written such that the fractional amplitude $a_m$ is normalized at the Einstein radius, and operates under the assumption that the reference EPL profile is near-isothermal (i.e., that it has a slope $t\sim 1$), which is a valid approximation for most lensing galaxies \citep[e.g.,][]{Auger2010, Shajib2021}.

\subsection{Curved arc basis}
\label{subsec:CAB_method}

Instead of relying on macromodel assumptions, \cite{Birrer_CAB} proposed to use a non-linear description of curved extended arcs, relying on a local basis formed with the eigenvectors of the lensing Jacobian. In practical strong lensing cases, the two eigenvectors are to a good approximation radial and tangential with respect to the center of the galaxy or galaxy cluster (see Figure~7 from \cite{Birrer_CAB}), so we write them $(\hat{e}_{\rm rad}, \hat{e}_{\rm tan})$, and have, by definition:

\begin{equation}
\begin{split}
    \mathbf{A}\cdot\hat{e}_{\rm rad}&= \lambda_{\rm rad}^{-1}\hat{e}_{\rm rad}\\
    \mathbf{A}\cdot\hat{e}_{\rm tan}&= \lambda_{\rm tan}^{-1}\hat{e}_{\rm tan}
\end{split}
\label{eq:eigenvalues_def}
\end{equation}
where the inverse eigenvalue $\lambda_{\rm rad}$ (respectively, $\lambda_{\rm tan}$) gives the stretch factor of the source in the radial (respectively, tangential) direction. The magnification is then simply:
\begin{equation}
    \mu = \frac{1}{\det(\mathbf{A})}  = \lambda_{\rm rad}\lambda_{\rm tan}.
    \label{eq:mag_CAB}
\end{equation}

For a more complete characterization of the local lensing field, one can introduce differentials along the eigenvectors:

\begin{equation}
    \partial_{\rm rad} \equiv \frac{\partial\ }{\partial\hat{e}_{\rm rad}}, \quad \partial_{\rm tan} \equiv \frac{\partial\ }{\partial\hat{e}_{\rm tan}}.
\end{equation}
For instance, we will consider the differentials of the eigenvalues along these directions ($\partial_i\lambda_j \text{ with } i,j = {\rm rad, tan}$). In addition, choosing a reference direction $\hat{e}_0$, a scalar angle $\phi_{\rm tan}$ giving the tangential eigenvector direction can be defined by:
\begin{equation}
    \cos \phi_{\rm tan}  = \hat{e}_{\rm tan} \cdot\hat{e}_0
\end{equation}
such that its differential along the tangential eigenvector is a measure of the tangential curvature
\begin{equation}
\begin{split}
    s_{\rm tan} \equiv \partial_{\rm tan}\phi_{\rm tan},
\end{split}
\label{eq:local_curv_def}
\end{equation}
independently of the choice for $\hat{e}_0$.

For the moment, Eqs.~(\ref{eq:eigenvalues_def})-(\ref{eq:local_curv_def}) are pointwise definitions (in particular, the gradients are written in the infinitesimal differential limit), but they can be extrapolated to provide the basis for a powerful non-linear parametrization of extended lensed arcs around a location of interest. This allows one to represent a great variety of curved arcs without the need of an explicit global deflector model \citep{Birrer_CAB}. 

\begin{figure}
    \centering
    \includegraphics[width=0.9\linewidth]{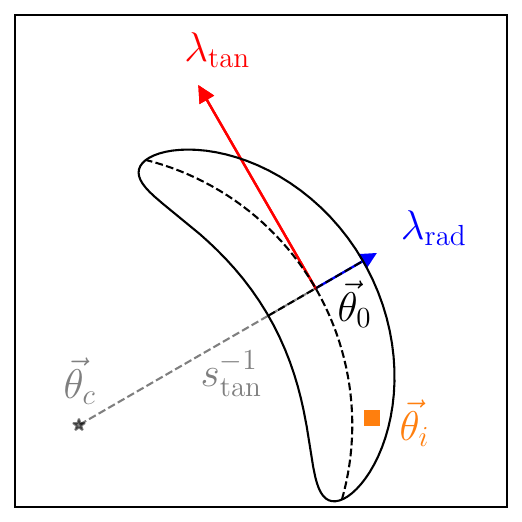}
    \caption{Illustration of the CAB formalism: a circular source is distorted into an arc, such that the lensing Jacobian at the center $\vec{\theta}_0$ of the arc has the eigenvectors shown by the red and blue arrows, with corresponding eigenvalues $\lambda_{\rm tan}$ and $\lambda_{\rm rad}$. Here, we present the simplest CAB model from Eq.~(\ref{eq:CAB_SIS+MST}): the arc lies on a circle centered on $\vec{\theta}_c$ with constant curvature radius $s_{\rm tan}^{-1}$, along which the eigenvalues are constant. The orange box depicts a generic pixel in the lensed arc, centered on $\vec{\theta}_i$ (used in a calculation in Appendix~\ref{App:SNR/I}).}
    \label{fig:CAB_illustration}
\end{figure}

CAB models exist in several versions, depending on the assumptions that one makes to extrapolate the local lensing properties. The simplest CAB description only relies on $6$ parameters: the two position coordinates of a reference location $\vec{\theta}_0$ (the center of the lensed arc) and the values of $\phi_{\rm tan}, \lambda_{\rm tan},\lambda_{\rm rad}$, and $s_{\rm tan}$ at that location. Then, making the following set of assumptions :
\begin{enumerate}
    \item the tangential curvature $s_{\rm tan}$ is constant, such that $\hat{e}_{\rm tan}$ describes a circle of radius $s_{\rm tan}^{-1}$,
    \item the eigenvalues $\lambda_{\rm tan},\lambda_{\rm rad}$ remain constant along that circle,
    \item there is no deflection shift at the center of the arc ($\vec{\alpha}(\vec{\theta}_0) = 0$),
\end{enumerate} 
and using the fact that the deflection field should be curl-free (i.e., $\frac{\partial\alpha_i}{\partial \theta_j}=\frac{\partial\alpha_j}{\partial \theta_i}$) if it derives from a lensing potential (see Section~\ref{subsec:grav_lensing_formalism} and Eq.~(\ref{eq:Jacobian_def}) in particular), we can write the following deflector model \citep{Birrer_CAB}:

\begin{equation}
\begin{split}
    &\vec{\alpha}(\vec{\theta}) = \lambda_{\rm rad}^{-1} \left[ \vec{\alpha}_{\rm SIS}(\vec{\theta}) - \vec{\alpha}_{\rm SIS}(\vec{\theta}_0) \right] + \left(1-\lambda_{\rm rad}^{-1}\right)\left(\vec{\theta}-\vec{\theta}_0\right)
     \\
    & \text{ where } \vec{\alpha}_{\rm SIS}(\vec{\theta}) = \theta_E^{\rm SIS} \frac{\vec{\theta}-\vec{\theta}_c}{\|\vec{\theta}-\vec{\theta}_c\|}, \quad \theta_E^{\rm SIS} = s_{\rm tan}^{-1} \left( 1 - \frac{\lambda_{\rm rad}}{\lambda_{\rm tan}}\right),
\end{split}
\label{eq:CAB_SIS+MST}
\end{equation}
and $\vec{\theta}_c = \vec{\theta}_0 - s_{\rm tan}^{-1} \hat{e}_{\rm rad}$ is the center of curvature. Eq.~(\ref{eq:CAB_SIS+MST}) is equivalent to the deflection field of a singular isothermal sphere \citep[SIS; e.g.,][]{Bartelmann2010} with Einstein radius $\theta_E^{\rm SIS}$, plus a mass sheet transformation \citep[MST ;][]{Falco1985, Gorenstein1988} with factor $\lambda_{\rm rad}^{-1}$, with the difference that it is intended to be a local description around a lensed arc, not a global model of the lensing field. This specific simple CAB model has previously been used to probe DM substructure in cluster-scale lenses \citep{Sengul2023, Ephremidze2025}. We provide an illustration of the CAB formalism in Figure~\ref{fig:CAB_illustration}, by applying the deflector model of Eq.~(\ref{eq:CAB_SIS+MST}) to a circular source, generating an arc with a shape that is fully characterized by the eigenvectors and tangential curvature at its center.

In the case of quadruply imaged quasars (with galaxy-scale lenses), if the quasar host galaxy is bright enough, each point-like quasar image is associated with a luminous extended arc. By modeling the arcs with independent CAB models, one can in theory predict the magnifications at the quasar image locations with “local” constraints on the lensing properties. The model described by Eq.~(\ref{eq:CAB_SIS+MST}), however, cannot account for asymmetric distortions in the lensed arc, for instance due to ellipticity in the lens mass distribution, or to external shear. These two effects can be incorporated by considering an extension to the simplest model \citep{Birrer_CAB}: we can relax assumption~(ii) and assume that the tangential stretch has a non-zero derivative $\partial_{\rm tan}\lambda_{\rm tan}$, evaluated at $\vec{\theta}_0$. Then, to satisfy the desired local properties while setting the next leading order to zero, a convenient way to write the deflector model is:
\begin{equation}
\begin{split}
    &\vec{\alpha}(\vec{\theta}) = \lambda_{\rm rad}^{-1} \left[ \vec{\alpha}_{\rm SIE}(\vec{\theta}) - \vec{\alpha}_{\rm SIE}(\vec{\theta}_0) \right] + \left(1-\lambda_{\rm rad}^{-1}\right)\left(\vec{\theta}-\vec{\theta}_0\right)
\end{split}
\label{eq:CAB_SIE+MST}
\end{equation}
where $\vec{\alpha}_{\rm SIE}(\vec{\theta})$ corresponds to the deflection field of a SIE \citep[e.g.,][]{Kassiola1993, Kormann1994,Schneider2006}, centered at $\vec{\theta}_c = \vec{\theta}_0 - s_{\rm tan}^{-1} \hat{e}_{\rm rad}$, and with Einstein radius $\theta_E$, semi-major axis orientation $\phi_{\rm SIE}$, and axis ratio $q$ given by:
\begin{equation}
    \begin{split}
        \theta_E &= s_{\rm tan}^{-1} \left( 1 - \frac{\lambda_{\rm rad}}{\lambda_{\rm tan}}\right) \sqrt{\frac{1+q^2}{2q}},\\
        \phi_{\rm SIE} &=  \phi_{\rm tan}+{\rm sgn}(\partial_{\rm tan}\lambda_{\rm tan})\frac{\pi}{4}, \\
        \text{and } q &= \sqrt{(1-\varepsilon)/(1+\varepsilon)} \text{ with }
        \varepsilon = \left| \frac{\partial_{\rm tan}\lambda_{\rm tan}}{s_{\rm tan}\cdot\lambda_{\rm tan}}  \left( 1 - \frac{\lambda_{\rm tan}}{\lambda_{\rm rad}}\right)^{-1} \right|.\\
    \end{split}
    \label{eq:CAB_SIE+MST_params}
\end{equation}

This extended CAB parametrization is equivalent to a \textit{local} SIE+MST model - adding freedom to one of the differentials ($\partial_{\rm tan}\lambda_{\rm tan} \neq 0$) allows more flexibility in the description. Relaxing other assumptions about the non-local extrapolation (e.g., allowing other differentials to be non-zero) will yield even more flexible models for lensed arcs. For instance, a generalization of the CAB formalism from \cite{Birrer_CAB} has been proposed by \cite{Sengul2025} in the context of cluster lenses. For galaxy-scale lenses, the arcs cover smaller angular scales and the mass/deflection field is \textit{a priori} less complex than for clusters, so the data does not require the same level of flexibility. In the following, we will show that the CAB model associated with Eqs.~(\ref{eq:CAB_SIE+MST}) and (\ref{eq:CAB_SIE+MST_params}) can already provide robust flux-ratio predictions for quadruply imaged quasars with extended arcs.

\section{Application to flux-ratio prediction} \label{sec:test}

In this Section, we describe how we implemented a CAB-based approach and investigated its capabilities in the context of flux-ratio prediction and anomaly detection in quadruply imaged quasars, testing its performance in comparison with the prevalent macromodel-based one. In Section~\ref{subsec:mock_obs}, we specify how we created simulated data sets with which to perform our tests, discussing, respectively, the mock lens (Section~\ref{subsubsec:mock_lens}), mock source (Section~\ref{subsubsec:mock_source}), and data quality (Section~\ref{subsubsec:mock_data_quality}). In Section~\ref{subsec:model_setup}, we describe how we modeled these mock images: we start with the shared source model (Section~\ref{subsubsec:source_model}), then present a standard lens macromodel approach for reference (Section~\ref{subsubsec:macromodel_fit}), and lastly, we detail our CAB-based flux-ratio prediction technique (Section~\ref{subsubsec:CAB_fit}).

The mock data were all created and analyzed using the lens modeling open-source software package \texttt{lenstronomy} \footnote{\url{https://github.com/lenstronomy/lenstronomy}}  \citep{Birrer2015,lenstronomy, lenstronomy2}, version v1.12.4. Figures~\ref{fig:CAB_illustration}-\ref{fig:baseline_CAB_fit}, \ref{fig:with_mult_CAB_fit}, \ref{fig:with_mult_macromodel_fit}, and \ref{fig:with_DM_CAB_fit} are also generated using this package.

\subsection{Mock observations}
\label{subsec:mock_obs}

To test the CAB fitting method discussed in Section~\ref{subsec:CAB_method} and compare its performance with a macromodel-based approach, we 
used simulated imaging data. We note that fitting each image is relatively expensive computationally, requiring many hours of supercomputer time. Therefore, we decided to run a set of simulations that is large enough to provide an initial exploration of the parameter space in terms of model complexity, lensing geometry, and signal-to-noise ratio (SNR), while leaving for future work the modeling of the exact properties of a specific dataset. In practice, we simulated imaging data for $27$ mock quadruply imaged quasars with extended arcs, randomly sampling source and lens parameters with distributions informed by the known properties of these systems \citep[e.g.,][]{Auger2010, Oguri2010}. We considered three cases for the mock lens:
\begin{enumerate}
    \item a “baseline” case, where the true lens properties follow a simple EPL+shear parametrization.
    \item a “complex” case, where, in addition to the EPL+shear profile, the lens mass distribution possesses azimuthal complexity in the form of elliptical multipole perturbations to the EPL profile.
    \item a “subhalo” case, where, in addition to the EPL+shear profile, a small DM halo is placed near one of the quasar images, in order to produce a significant flux-ratio anomaly.
\end{enumerate}

For each case, $9$ mock lenses were generated: specifically, 3 per basic lens configuration (cross, cusp, and fold), with varying levels of SNR in the extended arcs (see Section~\ref{subsubsec:mock_data_quality} and Appendix~\ref{App:SNR/I} for more detail).

\subsubsection{Mock lens parameters}
\label{subsubsec:mock_lens}
For all three cases, the true parameters of the EPL+shear were sampled according to the distributions listed in Table~\ref{tab:mock_params}. For case (ii), we added $m = 1$, $m = 3$, and $m = 4$ elliptical multipole terms with realistic directions and amplitudes, informed by observations \citep[e.g.,][]{Bender1988,Hao2006}. These choices are summarized in Table~\ref{tab:mock_params} ; in particular, to ensure enough azimuthal complexity relative to the baseline case, we fixed $a_4=\pm 1.5\%$ and $\varphi_4=0$ (i.e., we imposed significant diskyness/boxyness) since the $m=4$ order is expected to have a dominant effect, and to be aligned with the ellipse \citep[e.g.,][]{Bender1988, Hao2006, Oh2024}.

For case (iii), we employed the open-source package \texttt{pyHalo}\footnote{\url{https://github.com/dangilman/pyHalo}} \citep{Gilman2020}, designed to generate populations of DM substructure, to create individual subhalos. For simplicity, we placed the perturbing halo at the same redshift as the main lensing galaxy ($z_h = z_{\rm lens}=0.6$) and assumed that the 3D density followed a Navarro-Frank-White (NFW) profile \citep{NFW_1997}:  
\begin{equation}
    \rho_{\rm NFW}(r_{\rm 3D}) = \frac{\rho_s}{(r_{\rm 3D}/R_s)(1+r_{\rm 3D}/R_s)^2}
\end{equation}
where $\rho_s$ and $R_s$ are the characteristic density and scale radius, respectively. To generate physically realistic subhalos, we expressed $\rho_s$ and $R_s$ as a function of two other parameters: the mass $M_{200}$ and concentration $c_{\rm 200}$, implicitly defined by:
\begin{equation}
    \begin{split}
        M_{\rm 200} &\Big/ \left(\frac{4\pi}{3}R_{200}^3 \right) =  200 \cdot \rho_{\rm crit}(z_h) \\
         M_{\rm 200} &= 4\pi\rho_s R_s^3 \left[ \log(1+c_{\rm 200})-\frac{c_{\rm 200}}{1+c_{\rm 200}}\right]\\
         c_{\rm 200} &= R_{\rm 200}/R_s \\
    \end{split}
\end{equation}
where $\rho_{\rm crit}(z)$ is the critical density of the universe at redshift $z$, assuming a standard $\Lambda$CDM cosmology \citep{Planck2020}. To test whether CAB models would absorb the image perturbations from a DM halo in a variety of setups, the mass $M_{200}$ of the DM subhalo was drawn randomly from a log-uniform distribution in the $10^7-10^9M_\odot$ range. For $c_{200}$, we assumed CDM-like halos and used the concentration-mass relation from \cite{DiemerJoyce2019} with a scatter of $0.2$ dex. The image position of the subhalo was sampled uniformly in a disk of radius $\theta_{10}$ centered around one of the quasar images, where $\theta_{10}$ is defined by $\left|\log|\mu_{\rm NFW}(\|\vec{\theta}\|=\theta_{10})|-1\right|= 0.1$, i.e., as the radius at which the additional magnification from the NFW profile is $\sim10\%$, such that the subhalo would produce a significant flux-ratio anomaly ($\gtrsim 10\%$) in the mock image.

\addtolength{\tabcolsep}{2.5pt}
\begin{table}
\caption{\label{tab:mock_params} Distributions used to randomly sample true lens parameters for the mock imaging data. The elliptical multipoles are only added in case (ii). }
\begin{tabular}{lcr}
\hline
\textrm{Profile$^*$}&
\textrm{Parameter name}&
\textrm{Sampling distribution$^{**}$}\\
  \hline  \hline
EPL &  $\theta_E $ & $=3^{\prime\prime}$ \\
 & $t$ & $\mathcal{N}(1, 0.1)$ \\
 & $q$ & $\mathcal{U}(0.5, 0.95)$  \\
 & $\phi_{\rm EPL}$& $\mathcal{U}(-\pi, \pi)$ \\
 \hline
External shear & $\gamma_{\rm ext}$ & $\mathcal{U}(0, 0.15)$  \\
 & $\phi_{\rm ext}$ & $\mathcal{U}(-\pi, \pi)$ \\ 
  \hline \hline
Elliptical multipoles  &  &  \\
  \quad $m=1$ & $a_1$ & $\mathcal{N}(0, 0.005)$  \\
  & $\varphi_1$ & $\mathcal{U}(-\pi/2, \pi/2)$ \\ 
  \quad $m=3$ & $a_3$ & $\mathcal{N}(0, 0.005)$ \\
 & $\varphi_3$ & $\mathcal{U}(-\pi/6, \pi/6)$  \\ 
 \quad $m=4$ & $a_4$ & $=\pm 0.015$\\
 & $\varphi_4$ & $=0$ \\
 \hline
\end{tabular}\\
\footnotesize{$^*$The center of the lens profile is always assumed to be $(0,0)$.\\
$^{**}$We write $\mathcal{U}(a, b)$ for a uniform distribution on $[a,b]$ and $\mathcal{N}(\mu, \sigma)$ for a normal distribution.}
\vspace{-0.1cm}
\end{table}
\addtolength{\tabcolsep}{-2.5pt}

\vspace{-0.2cm}
\subsubsection{Mock source}
\label{subsubsec:mock_source}
\vspace{-0.1cm}

In order to create realistic extended lensed arcs, we used imaging data from the Cosmic Evolution Survey (COSMOS) \citep[e.g.,][]{Koekemoer2007} in the Hubble Space Telescope (HST) F814W filter to represent the surface light of the quasar host galaxy. Using the software package \texttt{paltas}  \citep{Wagner-Carena2023}, we selected $\sim 30$ galaxies with a variety of morphologies in the COSMOS/GREAT3 23.5 magnitude catalog \citep{COSMOS_catalog}, and randomly picked one for each simulated image (see Figure~\ref{fig:mock_example} for an example). This source was placed at a redshift $z_{\rm source}=1.9$ \citep[typical for lensed quasars,][]{Oguri2010}, with an angular position in the source plane $\vec{\beta}_s$ sampled inside the inner caustic of the mock lens in order to produce four distinct images. 

For the lensed quasar itself, we employed a simple point source and assumed a joint position with the center of the light profile representing its host galaxy. In the subhalo case, the magnification from small-scale structure is sensitive to the finite size of the region responsible for the background quasar's emission \citep[e.g.,][]{Dobler2006, Amara2006, Gilman2020}. We therefore assumed that flux ratios are “measured” with the quasar warm dust region, which is probed by the most recent James Webb Space Telescope (JWST) observations \citep{Nierenberg2024, Keeley2024} and spans typical scales of $\sim1-10$ pc. Specifically, we modeled the background source as a circular Gaussian with a FWHM of $5$ pc and, for each lensed image, we calculated magnifications by ray-tracing through a region tracking its surface brightness, then comparing the integrated fluxes in the source and image planes. Because the astrometric displacement of the quasar images due to the presence of the subhalo is typically negligible ($\lesssim 0.1$ mas) compared to the size of the ray-tracing region, this also allows us to directly compare the flux ratios with and without the subhalo.

In real data, there would be an additional light contribution from the main deflector galaxy. For simplicity, we did not include this component in our simulated data, effectively considering that the lens light would be sufficiently well-modeled to accurately remove its contribution at the location of the lensed arcs.

\subsubsection{Observing conditions}
\label{subsubsec:mock_data_quality}

Archival HST imaging data exists for most of the known quadruply imaged quasar systems \citep[e.g.,][]{Schmidt2023} - to generate and model the mock images, we therefore assumed observing specifications similar to HST, with pixel size $\delta_{\rm pix}=0.05 ^{\prime\prime}$, rms background noise $\sigma_{\rm bkg}=0.0058$ photons/s/pix, and exposure time $ t_{\rm exp} = 1428$s \citep[e.g.,][]{Shajib2019}. We employed a Gaussian point spread function (PSF) with width $\sigma_{\rm PSF}= 80$ mas (${\rm FWHM}\approx188$ mas), and added a combination of two sources of noise: a background component characterized by $\sigma_{\rm bkg}$, and photon shot noise described by Poisson statistics. Naturally, the results obtained here provide a generic guide: when applied to real data, one will have to match exactly the observing configuration.

\begin{figure*}
    \centering
    \includegraphics[width=0.97\textwidth]{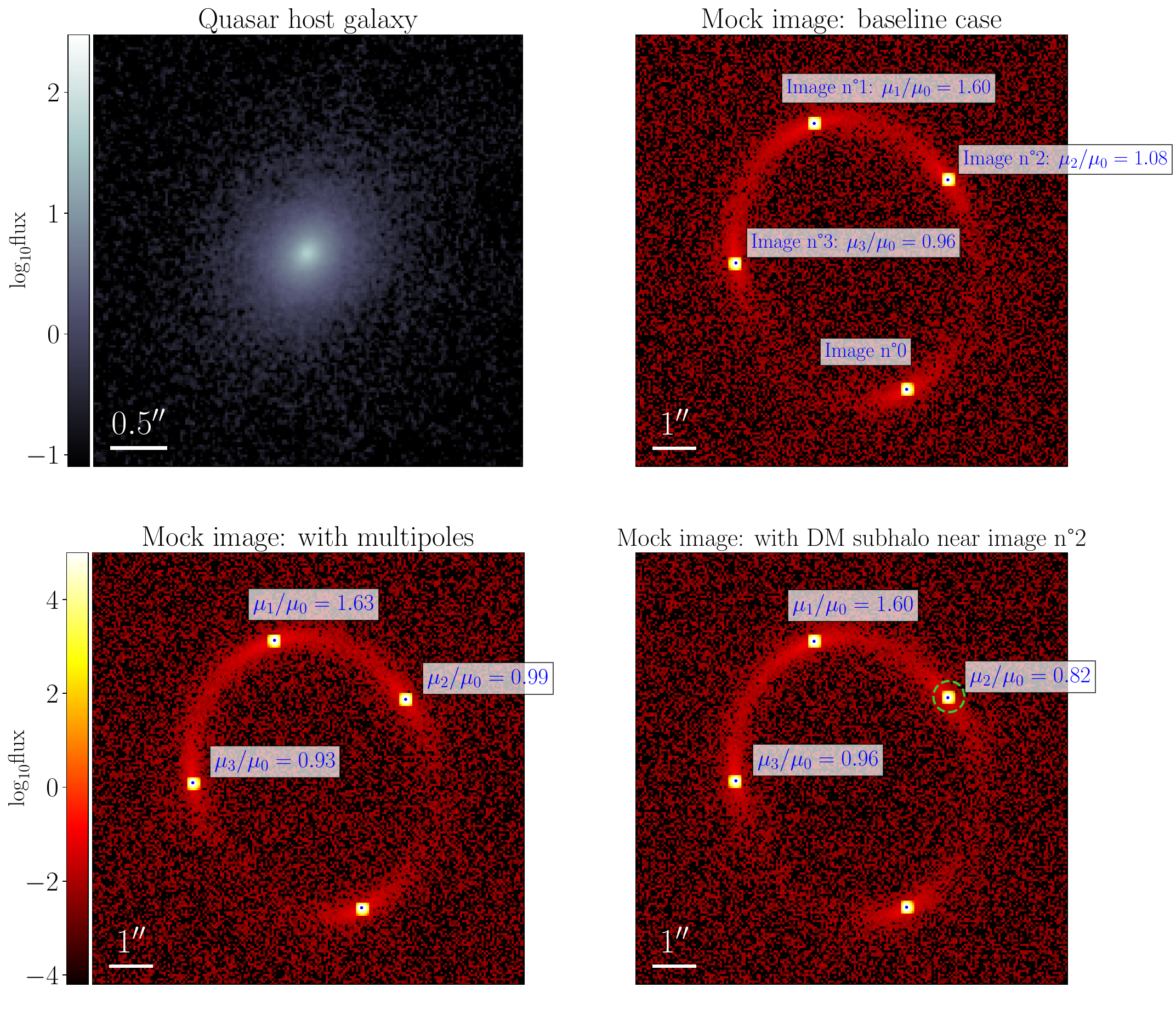}
    \caption{Illustrative examples of mock imaging data. Top left: surface photometry of a mock quasar host galaxy, drawn from the COSMOS catalog. Top right: mock imaging data in the baseline case, with a simple EPL+shear lens. The quasar images are labeled arbitrarily, with flux/magnification ratios reported relative to image n°0. Bottom left: mock imaging data in the complex case, with the same EPL+shear parameters as the baseline case but with added multipole perturbations (in particular $a_4=1.5\%$ and $\varphi_4=0$ to simulate diskyness). Bottom right: mock imaging data in the subhalo case, with the same EPL+shear parameters as the baseline case but with a DM subhalo near image n°2 causing a flux-ratio anomaly. The subhalo (with mass $M_{200}\approx 10^{8.7}M_\odot$ and concentration $c_{200}\approx11.2$) is located at the center of the light green circle, which has a radius of $2R_s$.}
    \label{fig:mock_example}
\end{figure*}

Because the CAB method relies on the extended lensed arcs, we adjusted the amplitude of the quasar host galaxy's light in order to simulate images that had sufficient lensing information in the arcs. This depends on the SNR of the pixels in the arc, but also its spatial extent since the CAB measures tangential/radial stretch. We discuss this in detail in Appendix~\ref{App:SNR/I} ; in the end, the choice of source amplitude resulted in images with $70<{\rm SNR}_{\rm arc}<200$ (where ${\rm SNR}_{\rm arc}$ is the total SNR in the lensed arcs, see Appendix~\ref{App:SNR/I}), with a large majority in the range $80<{\rm SNR}_{\rm arc}<130$. Figure~\ref{fig:mock_example} displays an example of mock imaging data for each of the three cases introduced in the introduction to Section~\ref{subsec:mock_obs}, generated on a $200\times200$ pixel grid. These examples are not part of the mock sample of $27$ lensed quasars as they share the same parameters for the mock source and for the EPL+shear part of the mock lens, in order to better illustrate the impact of the multipoles and of the DM subhalo on the imaging data and the flux ratios.

For the CAB fit (see Section~\ref{subsubsec:CAB_fit}), the lensing properties are modeled around each quasar image individually, so we defined four cutouts encompassing some of the lensed arc around the corresponding quasar image. We adopted cutouts with a default size of $40\times40$ pixels (i.e., $2^{\prime\prime}\times2^{\prime\prime}$), centered on the quasar images. For cusp and fold configurations, where several quasar images are close to one another, this choice would sometimes lead to some overlap between the cutouts, which would amount to counting the overlapping pixels several times in the likelihood. In order to avoid this, we iteratively tried to move the cutouts away from each other (with the quasar slightly off-center) and to shrink the cutout size by a few pixels, until the cutouts did not have a single pixel in common. Figures~\ref{fig:baseline_CAB_fit}, \ref{fig:with_mult_CAB_fit}, \ref{fig:with_mult_macromodel_fit}, and \ref{fig:with_DM_CAB_fit} display the cutouts used for each example mock image shown in this paper. 

Choosing an appropriate cutout size is important for the success of the CAB, as one needs to find a balance between the amount of information in the cutout, and the validity of the assumptions for the CAB deflection field (since the extrapolation of local properties are less accurate further away from the center of the arc). We discuss this extensively in Appendix~\ref{App:cutout}, where we show that excessively large cutouts can bias the recovered lens properties, and in Appendix~\ref{App:systematic_source}, where we show that the systematic uncertainty on the flux-ratio prediction is primarily driven by the physical size of the cutouts.

\subsection{Lens modeling: macromodel vs CAB}
\label{subsec:model_setup}

\subsubsection{Source model}
\label{subsubsec:source_model}

Whether the mock imaging data was fitted with a macromodel or CAB models, we employed the same default model for the source. Following a common method in lensed quasar modeling \citep[e.g.,][]{Schmidt2023, Keeley2024}, the quasar images were separately fitted in the image plane as four point sources, such that their positions and fluxes are not directly determined by the lens model. This ensures that the observed image fluxes (which might be perturbed by substructure) do not affect the smooth lens model. The fitted image positions, however, are used as constraints when sampling lens parameters - demanding that they map back to the same location in the source plane and thus effectively enforcing the lens equation \citep{Birrer2015, Gilman2020}. The quasar image positions are allowed to vary within their astrometric uncertainties, which we assumed to be $\sim 4$ mas based on recent assessments \citep{Nierenberg2024, Keeley2024}.

The light distribution of the quasar host galaxy in the source plane was modeled to first order by an elliptical Sérsic profile \citep{Sersic1963, Sersic1968, Graham2005} with index $0.5<n_s<8$.
To capture additional small-scale complexity expected in the COSMOS images, we also included shapelet components, implemented as a basis of Hermite-Gaussian functions \citep{Refregier2003,Birrer2015}, with maximum order $n_{\rm max}$ characterizing the level of detail in the reconstructed image. In a real data modeling scenario, one would need to optimize $n_{\rm max}$ to find the appropriate level of complexity in the source model ; for simplicity, we fixed $n_{\rm max}=5$ since this is typically enough degrees of freedom for real data \citep[e.g.,][]{Shajib2019, Keeley2024}, and we are working in an idealized setup.

\subsubsection{Macromodel-predicted flux ratios}
\label{subsubsec:macromodel_fit}

For reference, we first fitted the mock imaging data with a typical macromodel-based approach. In order to highlight the potential pitfalls of this method, we employed (unless stated explicitly) a simple EPL+shear model, regardless of the true mock lens properties. In particular, for case (ii), the elliptical multipoles used to generate the data were not included in the macromodel at first. 
The idea is that multipoles then represent a proxy for any unmodeled complexity in the lens mass distribution - in current DM substructure analyses with flux-ratio anomaly statistics \citep[e.g.,][]{Gilman2022, Keeley2024, Gilman2024}, multipole perturbations are included in the lens models, but there might be other deviations from EPL+shear (isodensity twists, ellipticity gradients, etc.) that are not captured.

To find the model-predicted flux ratios, we followed a standard \texttt{lenstronomy} sequential fitting procedure \citep{Birrer2015,lenstronomy,lenstronomy2}, first finding the non-linear model parameters maximizing the imaging likelihood with a Particle Swarm Optimization (PSO) algorithm ; then estimating the posterior probability distribution of the same parameters using Markov Chain Monte-Carlo (MCMC) sampling. We note that, in order to make a more direct comparison with the CAB fit (see Section~\ref{subsubsec:CAB_fit}), the imaging likelihood only includes pixels that are within the four cutouts described in Section~\ref{subsubsec:mock_data_quality}. By calculating the magnification at the image positions associated with the parameters in each MCMC sample, we then obtained posterior distributions for the macromodel-predicted flux ratios. We find that the model-predicted uncertainties are best represented by Gaussian distributions in logarithmic space, so we define the log magnifications and log flux-ratios:
\begin{equation}
\begin{split}
    \mathcal{M}_i \equiv & \log |\mu_i| \quad \text{ for } 0\leq i \leq3 ,\\
    \mathcal{F}_{ij} \equiv & \mathcal{M}_i-\mathcal{M}_j \text{ for } 0\leq i ,j\leq3,
\end{split}
\label{eq:log_mag_log_FR}
\end{equation}
and report results with these quantities - see Section~\ref{sec:results}. Specifically, for the figures, we only display 3 log flux-ratios (arbitrarily: $\mathcal{F}_{01}, \mathcal{F}_{02}$, and $\mathcal{F}_{03}$, where the labels bear no physical meaning) since the other values are not linearly independent (e.g., $\mathcal{F}_{12}=\mathcal{F}_{02}-\mathcal{F}_{01}$). We also avoid showing the (log-)magnifications since their values are degenerate with the intrinsic luminosity of the source.

\subsubsection{CAB model-predicted flux ratios}
\label{subsubsec:CAB_fit}

For the CAB fit, we chose four independent CAB models to represent the lensing properties in each cutout (i.e., around each quasar image) individually. Specifically, we used the tangentially varying deflector model from Eqs~(\ref{eq:CAB_SIE+MST}) and (\ref{eq:CAB_SIE+MST_params}), which allows $\partial_{\rm tan}\lambda_{\rm tan} \neq0$ for more flexibility. We note that the fits in each cutout were not completely independent (only the parameters describing the lensing properties), since the source was reconstructed jointly between the four CAB models.
Because the deflector model of Eq~(\ref{eq:CAB_SIE+MST}) is set up with the convention $\vec{\alpha}(\vec{\theta}_0)=0$, we also fitted a constant deflection shift $\vec{\alpha}_0$ independently for each cutout - because of the prismatic degeneracy \citep{Gorenstein1988}, these parameters have no impact on the lensing observables. We anchored the reference location of the CAB expansion ($\vec{\theta}_0$) at the fitted quasar image location, such that the magnification at that location can directly be calculated with Eq.~(\ref{eq:mag_CAB}), using the fitted values for $\lambda_{\rm tan} \text{ and } \lambda_{\rm rad} $. For the same reason as the macromodel case, we will report the results in terms of the log flux-ratios defined in Eq.~(\ref{eq:log_mag_log_FR}).

During the first step of the CAB fit, we used the same procedure as for the macromodel, running a PSO then a MCMC sampler. For systems with low SNR in the lensed arcs, however, the posterior distribution obtained with MCMC sampling for CAB-predicted values frequently displayed very large uncertainties (sometimes $\gtrsim30\%$ on flux ratios). This would make the detection of flux-ratio anomalies very difficult in practice, since it would both dominate the typical measurement uncertainties  \citep[$~2-10\%$,][]{Nierenberg2024, Keeley2024} and wash out any perturbation from substructure. Since this issue is driven by pixel-level noise in the image, we circumvented it by resampling the noise map several times\footnote{When applying this technique on real data, instead of resampling the exact noise map (which is not known \textit{a priori}), one can analogously resample the imaging data residuals after achieving an acceptable fit.}, then using bootstrapped estimators for the log flux-ratios and statistical uncertainties, which are more robust than values found for a single noise realization. In our analysis, we resampled the noise map $N_s=30$ times for each mock lens, running a PSO algorithm to find the best-fitting parameters for each noise realization. Then, the estimator for the log flux-ratios was chosen to be:
\begin{equation}
    \widehat{\mathcal{F}}_{ij} = \text{med} \left( \left\{\mathcal{F}^{[k]} \right\}_{1\leq k \leq N_s} \right)_{ij} \text{ for } 0 \leq i<j\leq3
    \label{eq:log_FR_estimator}
\end{equation}
where $\text{med}$ is the (6D) geometric median and $\mathcal{F}_{ij}^{[k]}$ are the CAB-predicted flux ratios given by the best-fitting parameters of the $k$-th noise realization. Due to computational limitations, we only resampled the noise map a small number of times, so we used the median instead of the mean in order to mitigate the impact of potential outliers, which can be observed in predicted flux-ratio space if the PSO fails to converge or finds a secondary peak in the likelihood.

To estimate the uncertainty in the CAB-predicted flux ratios, we considered both statistical and systematic errors. Since the statistical uncertainty is driven by the noise in the image, we used the expected variance of the estimator from Eq.~(\ref{eq:log_FR_estimator}), assuming that every noise realization had an uncertainty on $\mathcal{F}_{ij}^{[k]}$ similar to the one obtained with MCMC sampling. We estimated the systematic uncertainty by comparing the flux ratios predicted by the best-fitting CAB model to the true values in hundreds of mock lens images with very high SNR (${\rm SNR}_{\rm arc} \gtrsim 10^3$), simulated using the pipeline from Section~\ref{subsec:mock_obs}.  
A detailed discussion of the uncertainty budget for the CAB-predicted flux ratios is presented in Appendix~\ref{App:uncertainties}. The results shown in Section~\ref{sec:results} reflect the final uncertainties estimated using this method.

\section{Results} \label{sec:results}

\begin{figure*}
    \centering
    \includegraphics[width=0.94\textwidth]{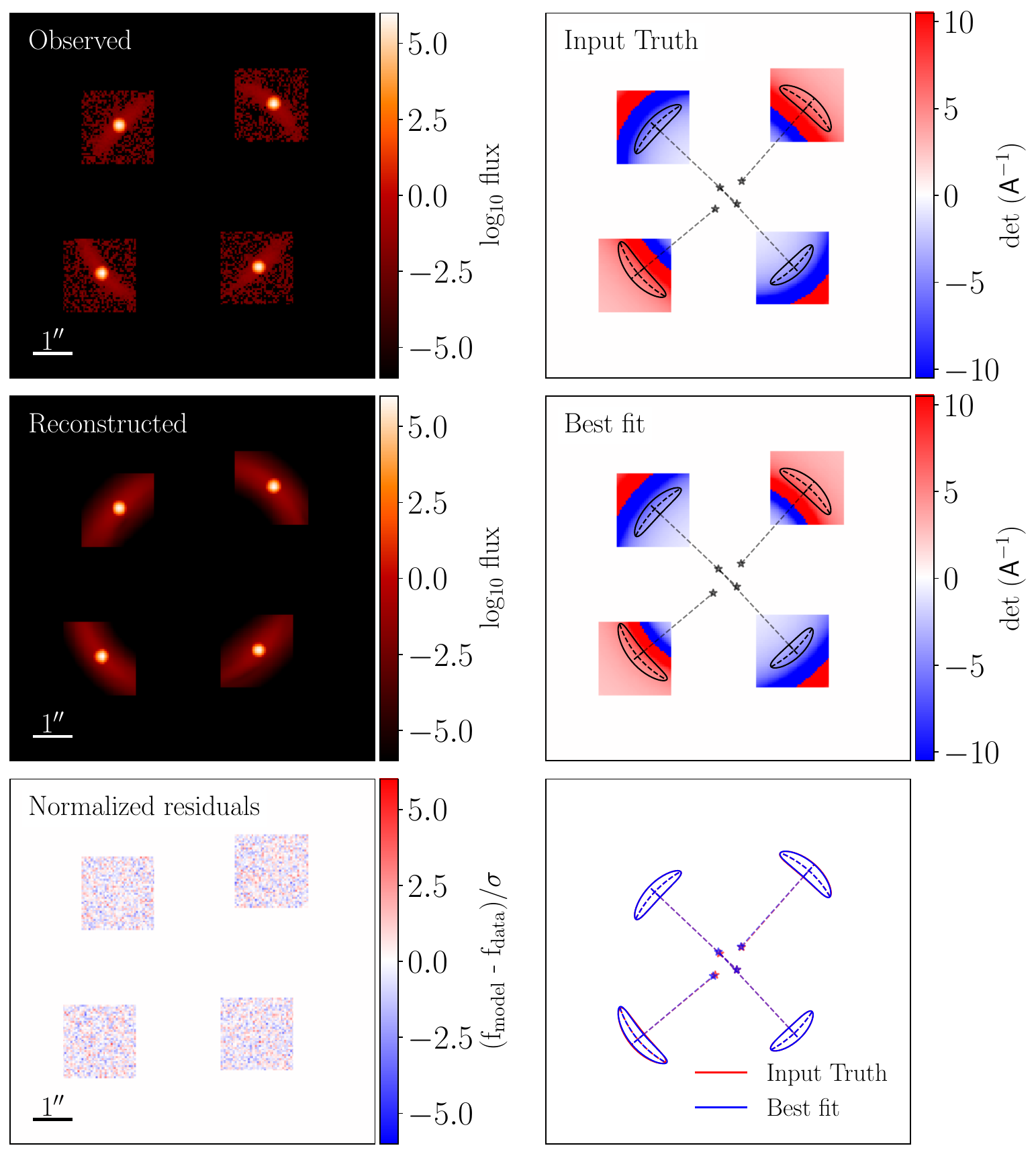}
    \caption{Example of CAB fit for a lensed system in the baseline case. This is a cross-like system with ${\rm SNR}_{\rm arc}\approx 93$. Top row: “true” quantities for one of the noise map realizations. Middle row: quantities under best-fitting CAB model for the same noise realization. Bottom row: comparison between true and best-fitting quantities.
    Left column: imaging data in the four cutouts. Right column: magnification maps in the four cutouts and visual representation of a curved arc defined by the local lensing properties at the quasar image locations (similar to Figure~\ref{fig:CAB_illustration}).}
    \label{fig:baseline_CAB_fit}
    \vspace{-0.2cm}
\end{figure*}

\begin{figure*}
    \centering
    \includegraphics[width=0.95\textwidth]{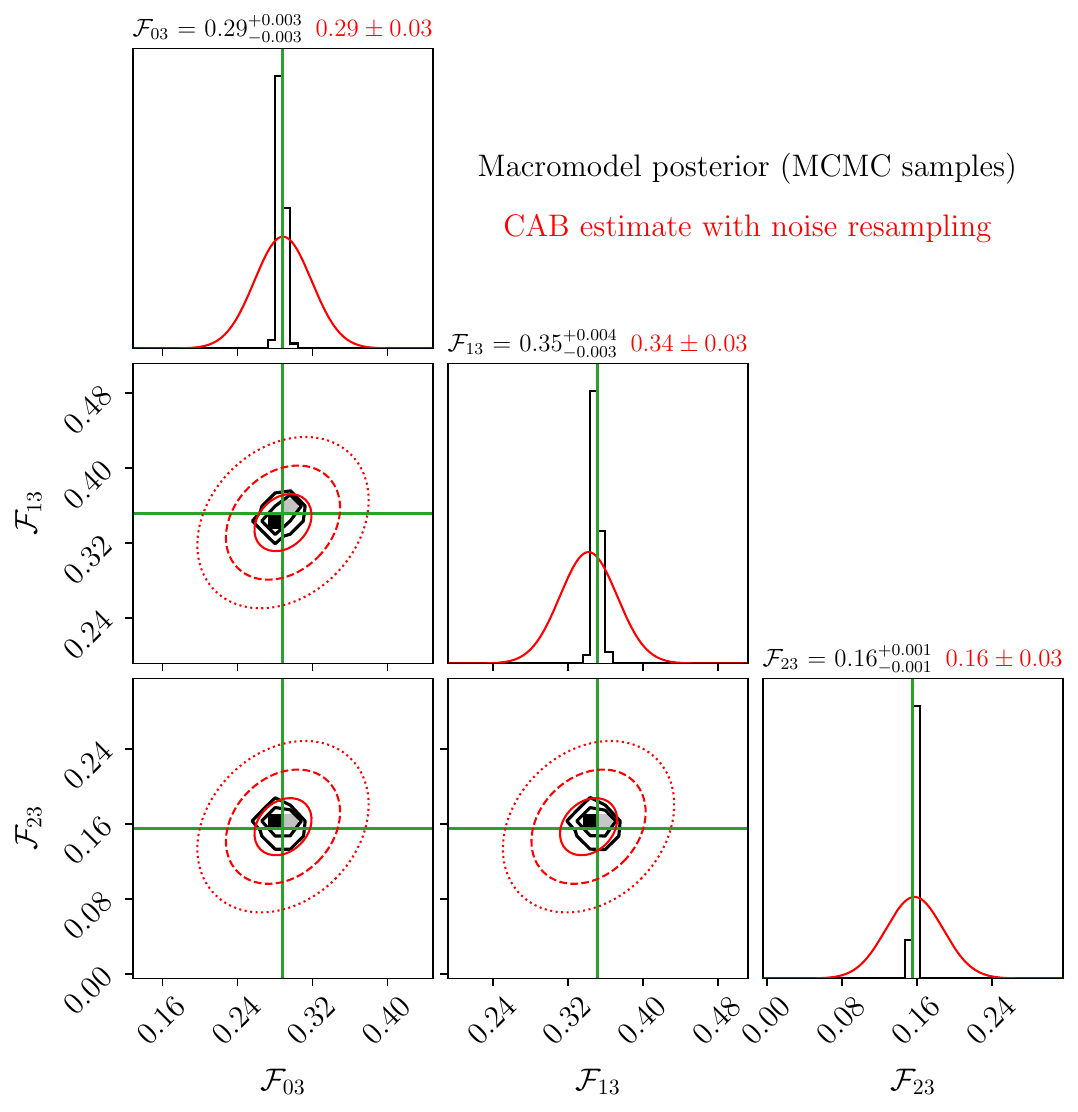}
    \caption{Posterior probability distribution of three of the (log) flux-ratios for the baseline case example shown in Fig.~\ref{fig:baseline_CAB_fit}, under the macromodel (in black) and the CAB model (in red). The macromodel posterior is directly estimated from the MCMC samples, and the reported quantity and lower/upper errors are the median and the $16/84\%$ quantiles, respectively.  The CAB posterior is a multivariate Gaussian with bootstrapped estimates for the mean and covariance, obtained by resampling the noise map (see Section~\ref{subsubsec:CAB_fit}). The green crosshairs show the true flux ratios, i.e., the magnification ratios from the true lensing field. }
    \label{fig:baseline_FR_prediction}
\end{figure*}

\begin{figure*}
    \centering
    \includegraphics[width=0.94\textwidth]{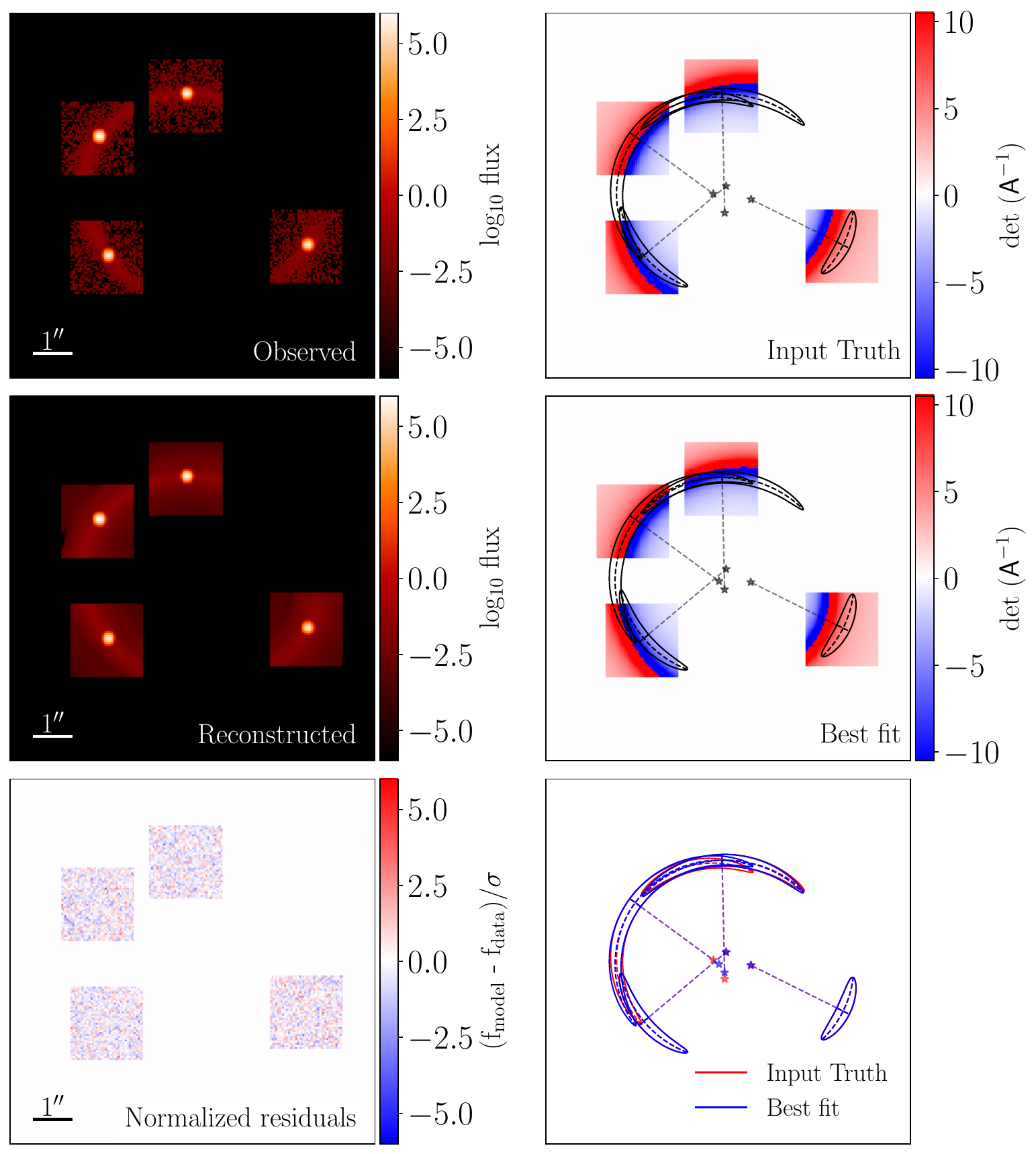}
    \caption{Same as Figure~\ref{fig:baseline_CAB_fit}, but for a lensed system in the complex case, which has large multipoles (in particular, $a_4=1.5\%$, i.e., significant diskyness in this case) added to the underlying lens mass distribution. This is a cusp-like system with ${\rm SNR}_{\rm arc}\approx 111$.}
    \vspace{-0.25cm}
    \label{fig:with_mult_CAB_fit}
\end{figure*}

\begin{figure*}
    \centering
    \includegraphics[width=0.95\textwidth]{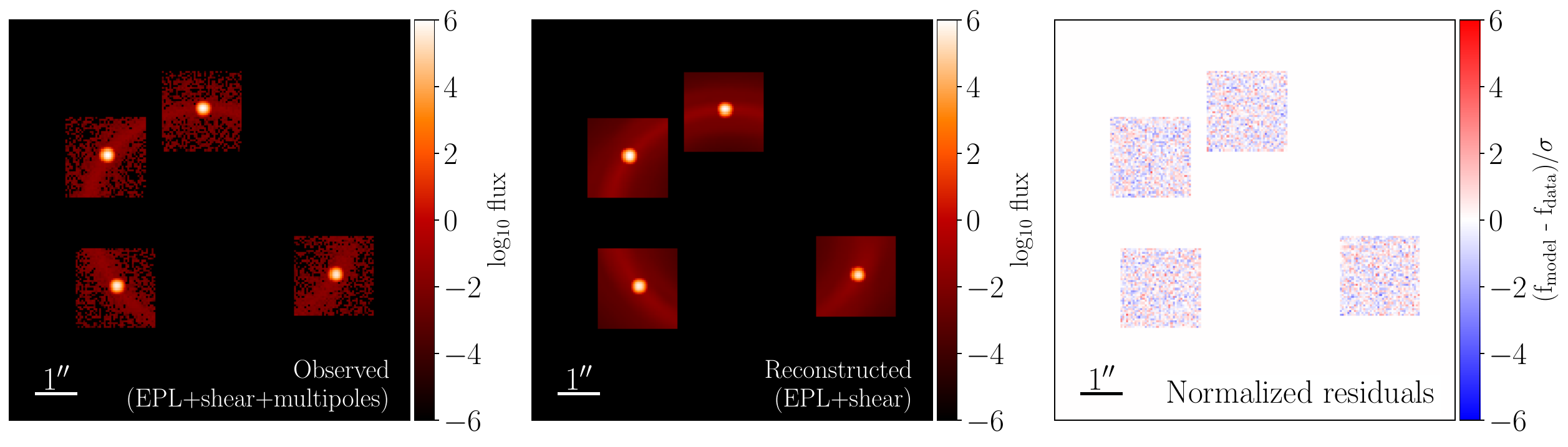}
        \caption{Imaging data reconstruction with the best-fitting oversimplistic macromodel (EPL+shear, i.e., without multipoles), for the same lensed system as Fig.~\ref{fig:with_mult_CAB_fit} (the mock data was generated with large multipoles, in this case a significantly disky lens).}
    \label{fig:with_mult_macromodel_fit}
    \vspace{-0.5cm}
\end{figure*}

\begin{figure*}
    \centering
    \includegraphics[width=0.9\textwidth]{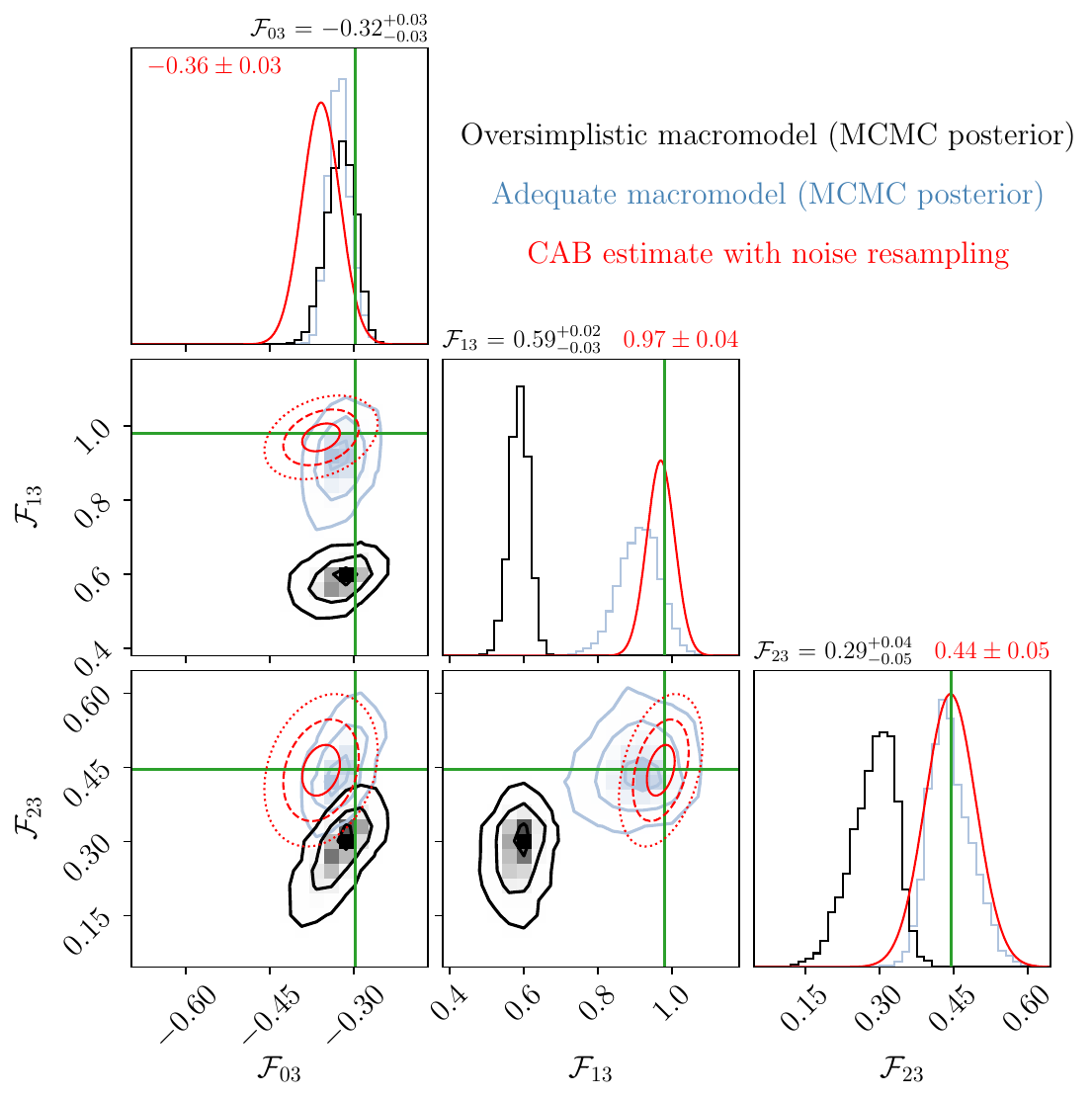}
    \caption{Same as Figure~\ref{fig:baseline_FR_prediction}, but for the lensed system from Figures~\ref{fig:with_mult_CAB_fit} and \ref{fig:with_mult_macromodel_fit}, which has large multipoles in the true lens mass distribution (in particular, significant diskyness). The black posterior corresponds to an EPL+shear only macromodel, which is does not capture this additional complexity. In light blue, we add the posterior distribution, estimated with MCMC samples, under a more adequate macromodel that includes multipoles (specifically, we added $m=1,m=3,\text{ and }m=4$ elliptical multipoles with free parameters to the EPL+shear lens mass model).}
    \vspace{-0.3cm}
    \label{fig:with_mult_FR_prediction}
\end{figure*}

\begin{figure*}
    \centering
    \includegraphics[width=0.94\textwidth]{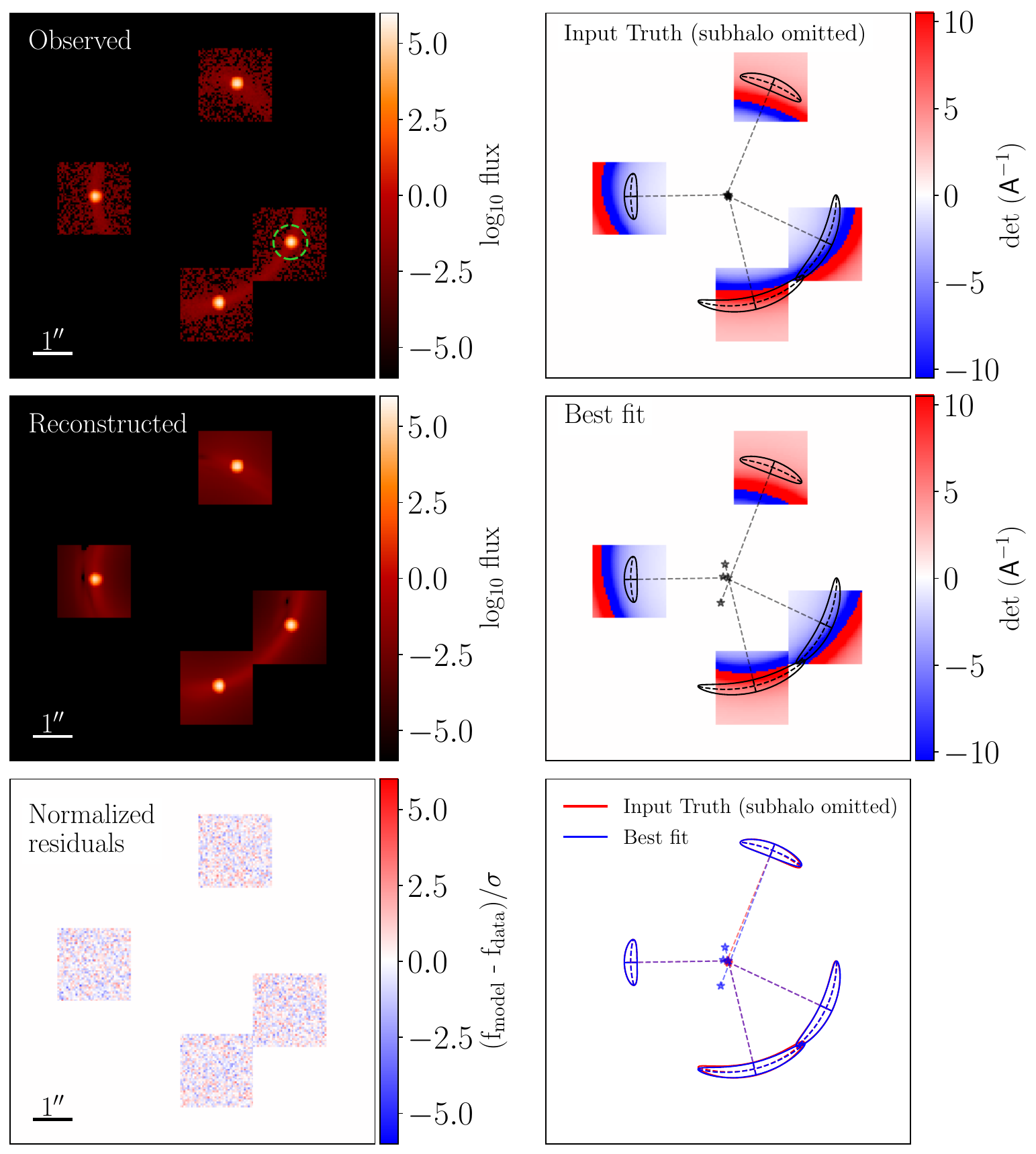}
    \caption{Same as Figures~\ref{fig:baseline_CAB_fit} and \ref{fig:with_mult_CAB_fit}, but for a lensed system in the subhalo case, i.e. with an individual DM subhalo injected near one of the images in order to produce a significant flux-ratio anomaly. The magnification map and representation of CAB quantities for the input truth (upper and lower right panels) do not include the subhalo contribution. This is a fold-like system with ${\rm SNR}_{\rm arc}\approx 94$. The subhalo (with mass $M_{200}\approx 10^{8.5}M_\odot$ and concentration $c_{200}\approx7.5$) was injected at the center of the light green circle (top left panel), which has a radius of $2R_s$.}
    \label{fig:with_DM_CAB_fit}
\end{figure*}

\begin{figure*}
    \centering
    \includegraphics[width=0.95\textwidth]{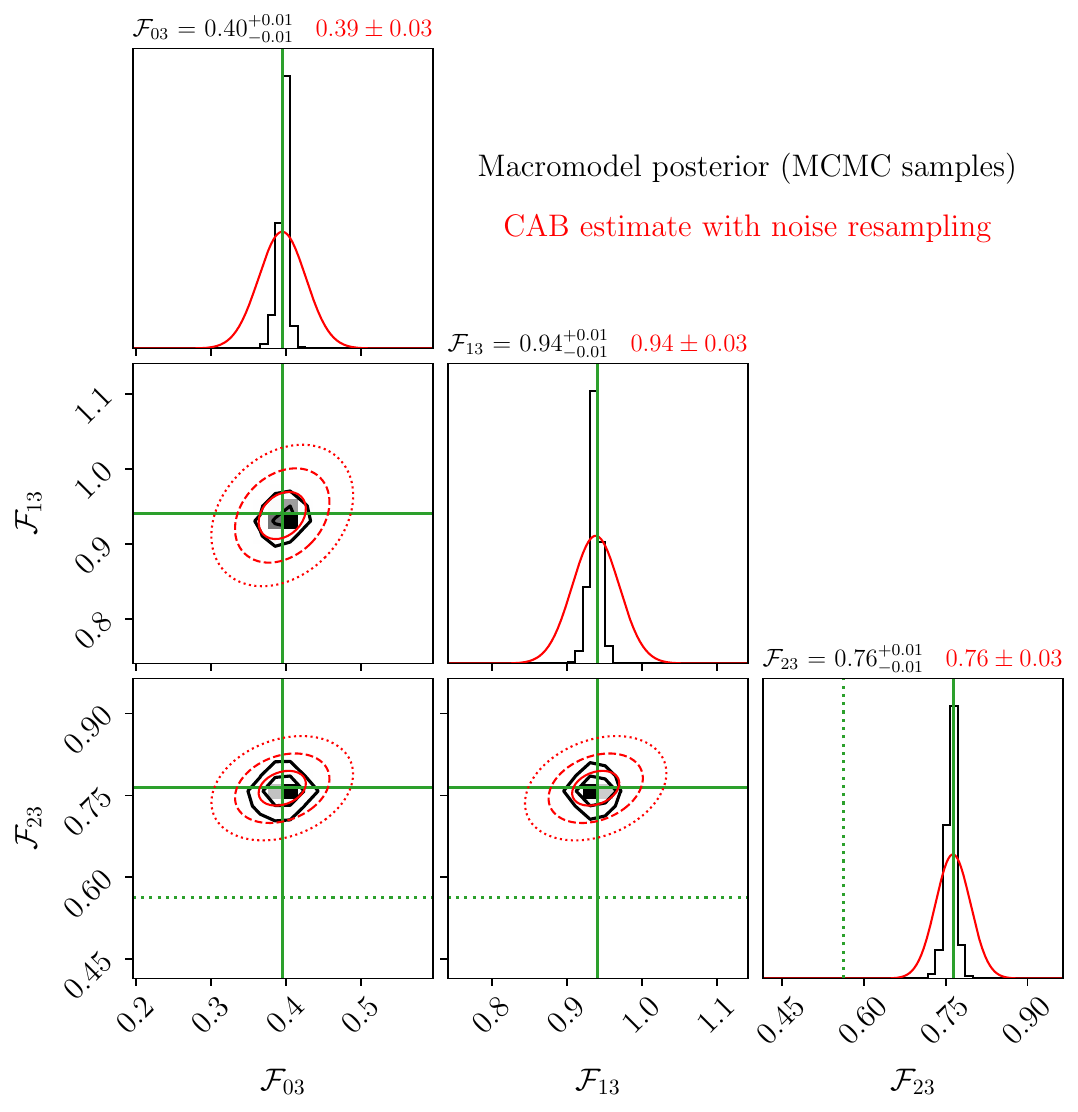}
    \caption{Same as Figure~\ref{fig:baseline_FR_prediction}, but for the lensed system from Figure~\ref{fig:with_DM_CAB_fit}, which has an anomaly-inducing DM subhalo near image n°2. The solid green crosshairs show what the flux ratios would be without the DM subhalo, i.e., the magnification ratios from the main deflector alone - we expect that an accurate smooth lens model will reproduce those values. The dashed green crosshairs correspond to the flux ratios that would actually be observed, including the effect of the subhalo: image n°2 alone is perturbed, such that only $\mathcal{F}_{23}$ is significantly impacted. Both the macromodel and the CAB model would detect a true positive flux-ratio anomaly in that case, i.e., a discrepancy between model prediction and observation that is caused directly by DM substructure.}
    \label{fig:with_DM_FR_prediction}
\end{figure*}

\begin{figure*}
    \centering
    \includegraphics[width=0.94\textwidth]{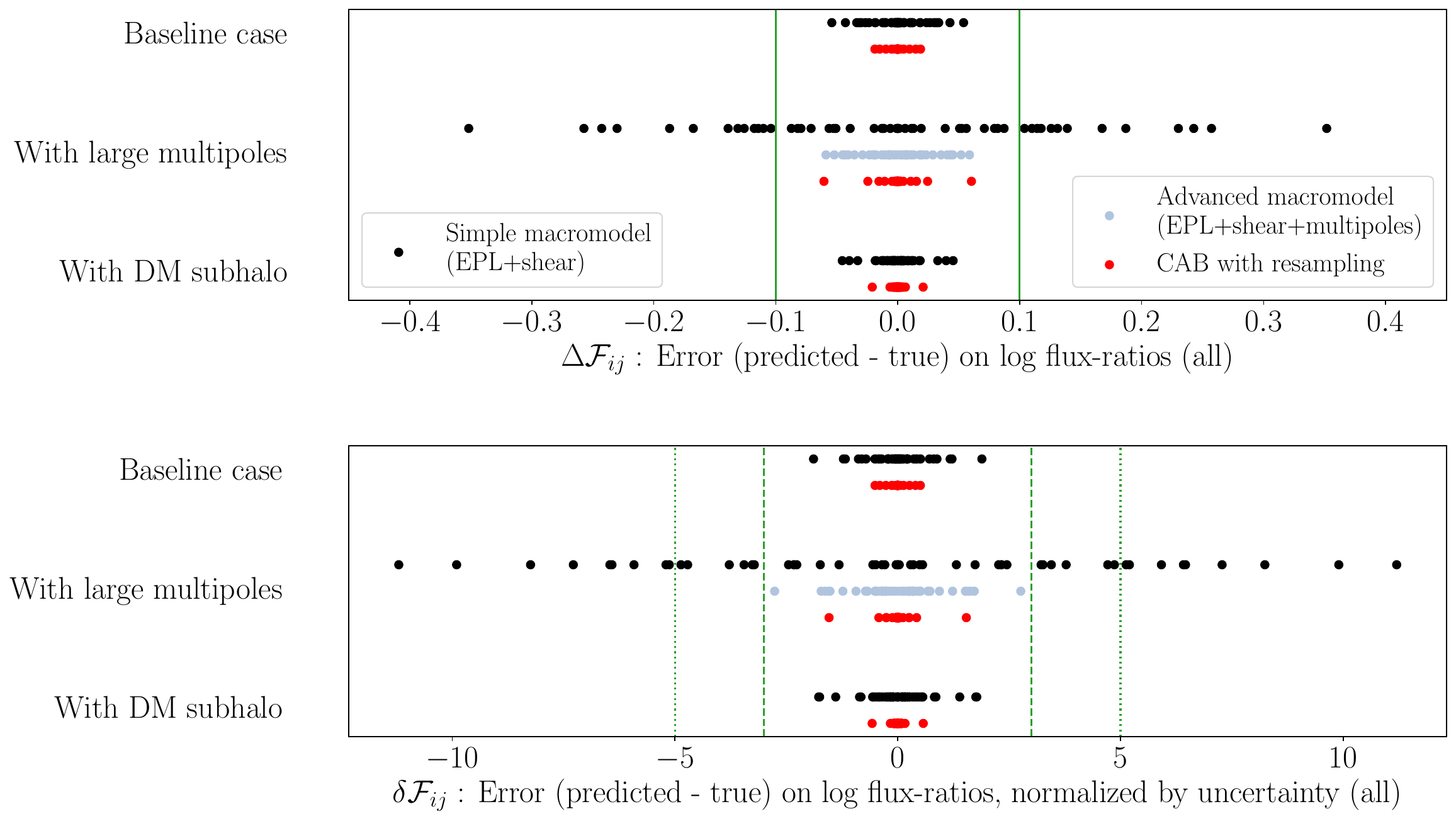}
    \caption{Top: distribution of $\Delta\mathcal{F}_{ij}$ (absolute prediction error on log flux-ratios, see Eq.~(\ref{eq:prediction_errors})), for all flux ratios in each mock lens case. The solid green lines correspond to a $\sim 10\%$ discrepancy between expected and true value. For the subhalo case, we show the difference between the model-predicted values and the values that would be observed without the subhalo. Bottom: distribution of $\delta\mathcal{F}_{ij}$ (prediction error normalized by a combination of model and measurement uncertainties, see Eq.~(\ref{eq:prediction_errors})), for all flux ratios in each mock lens case. The dashed and dotted green lines indicate a $3\sigma$ and $5\sigma$ discrepancy, respectively. }
    \vspace{-0.1cm}
    \label{fig:summary_FR_prediction}
\end{figure*}

In this section, we discuss the model-fitting and flux-ratio prediction performance of the EPL+shear macromodel and CAB model for the three mock lens cases. We exhibit one representative example for each: see Figures~\ref{fig:baseline_CAB_fit}-\ref{fig:baseline_FR_prediction} for the baseline case (Section~\ref{subsubsec:baseline_results}), Figures~\ref{fig:with_mult_CAB_fit}-\ref{fig:with_mult_FR_prediction} for the complex case (Section~\ref{subsubsec:with_mult_results}), and Figures~\ref{fig:with_DM_CAB_fit}-\ref{fig:with_DM_FR_prediction} for the subhalo case (Section~\ref{subsubsec:with_DM_results}). To summarize our results regarding flux-ratio prediction, we compute the following quantities for each mock lens:
\begin{equation}
\begin{split}
    \Delta\mathcal{F}_{ij} &= \mathcal{F}_{ij}^{\rm pred} - \mathcal{F}_{ij}^{\rm true} \\
     \delta\mathcal{F}_{ij} &= \frac{\mathcal{F}_{ij}^{\rm pred} - \mathcal{F}_{ij}^{\rm true}}{\sqrt{\sigma_{ij}^2+\sigma_{\rm meas}^2}}
     \label{eq:prediction_errors}
\end{split}
\end{equation}
where $\sigma_{\rm meas}$ is the expected measurement uncertainty \citep[we chose $\sigma_{\rm meas}=0.02$ based on recent JWST results,][]{Nierenberg2024, Keeley2024}, $\mathcal{F}_{ij}^{\rm true}$ is the log flux-ratio value that the lens model should predict (i.e., the magnification ratio from the main deflector, not including any subhalos), and $\mathcal{F}_{ij}^{\rm pred}, \sigma_{ij}$ are the model-predicted value and uncertainty (using either the CAB or the macromodel). Figure~\ref{fig:summary_FR_prediction} reports these absolute ($\Delta\mathcal{F}_{ij}$) and normalized ($\delta\mathcal{F}_{ij}$) prediction errors for the 6 image pairs $ \times$ 9 mock lenses $=$ 54  flux ratios in each mock lens case.

For the model uncertainties $\sigma_{ij}$, we used the half-width of the 50\% central credible interval in the marginalized distribution of MCMC samples (for the macromodel prediction), or the estimator discussed in Section~\ref{subsec:CAB_method} and Appendix~\ref{App:uncertainties}, which combines statistical and systematic uncertainty estimates (for the CAB prediction). Regardless of the mock lens case, by examining the distribution of errors on mocks with high SNR, we found that, for our choice of cutout, the systematic uncertainty on $\mathcal{F}_{ij}$ for the CAB estimate was typically $\sigma_s\sim 0.03$ (i.e., $\sim\text{3}\%$ uncertainty on the flux-ratio value). Detailed tests hint that this is due to the assumptions in the CAB deflector model, i.e., assumptions about how the extrapolation of local properties is performed (see Appendix~\ref{App:uncertainties} and \ref{App:systematic_source} for a detailed discussion).

\subsection{Baseline case: ideal macromodel}
\label{subsubsec:baseline_results}

We start by presenting results for the baseline case, i.e., in the absence of DM substructure, and when the macromodel used to fit the data is the same as the one used to generate the data (EPL+shear, without multipoles). For the $9$ mock lenses, we find, unsurprisingly, that both the EPL+shear macromodel and the CAB model are able to reproduce the imaging data down to noise level (see Figure~\ref{fig:baseline_CAB_fit} for an example of imaging data modeled with a CAB fit in the baseline case).

The log flux-ratios predicted by both methods are consistent with the true values, within their respective uncertainties (see Figure~\ref{fig:baseline_FR_prediction} for an example, and Figure~\ref{fig:summary_FR_prediction} for a summary of the flux-ratio prediction performance). The macromodel is, however, more precise, in the sense that the uncertainties obtained from the MCMC posterior of the macromodel fit are smaller than both the uncertainties from the MCMC posterior of the CAB fit for a single realization of the noise map, and the uncertainties estimated using the resampling method described in Section~\ref{subsubsec:CAB_fit}  (see Figure~\ref{fig:baseline_FR_prediction} for an example). This is expected since the CAB model relaxes some of the assumptions that are imposed in a macromodel fit (specifically, the global parametrization of the deflection field). 

For the estimated CAB uncertainties, if enough noise realizations are considered for the resampling, the statistical uncertainty can in theory be shrunk to arbitrarily small values. Even with our small number $N_s=30$ of resamples, the large uncertainty associated with a single noise realization can be reduced to levels similar to (and in most cases negligible compared to) the noise floor, i.e., the systematic uncertainty term. The final uncertainties on the CAB-predicted flux ratios are therefore typically $\sim \text{3-5}\%$, which is reasonable when compared to the expected measurement uncertainties for the flux ratios  \citep[$\sim2-10\%$,][]{Nierenberg2024, Keeley2024}.

\subsection{Complex case: false-positive flux-ratio anomalies}
\label{subsubsec:with_mult_results}

In this section, we discuss the impact of adding some complexity in the underlying lens mass distribution (in the form of multipole perturbations, including significant diskyness/boxyness, see Section~\ref{subsubsec:mock_lens}) that is not explicitly included in the models. Again, both the macromodel and CAB methods provide good fits to the imaging data for the 9 lenses in this “complex” case, with little to no visible structure in the normalized residuals - in Figures~\ref{fig:with_mult_CAB_fit} and \ref{fig:with_mult_macromodel_fit}, we show an example mock lens with its best-fitting CAB model and EPL+shear macromodel, respectively. This means that angular complexity in the mass can be compensated (to some extent) in the image fitting by adjusting some of the EPL+shear parameters \citep[in particular, it is often suspected that external shear is used as a “fudge factor” to correct for a lack of complexity in the mass models, e.g.,][]{Etherington2024} - which are then recovered with some systematic errors.
\newpage

As a consequence, we find in this case that the flux-ratio values predicted by the insufficiently complex macromodel are, in many instances, not statistically consistent with the true values. We show an example in Figure~\ref{fig:with_mult_FR_prediction}: the true flux ratio between images n°1 and 3 is $\gtrsim10\sigma$ away from the prediction of the EPL+shear macromodel - a $\sim 40\%$ flux-ratio anomaly would be inferred, despite the absence of DM substructure. This is consistent with the findings from previous studies \citep[e.g.,][]{EvansWitt2003, O'Riordan2024, Cohen2024, Nightingale2024}, which suggest that specific combinations of multipole perturbations can mimic the effect of a DM substructure, which motivates the inclusion of such perturbations in flux-ratio anomaly statistics analyses \citep[e.g.][]{Gilman2022,Gilman2024, Keeley2024}.

More specifically, 8 out of the 9 lensed systems simulated in the complex case display false-positive flux-ratio anomalies greater than 10\% when using the EPL+shear macromodel, with sometimes several anomalous flux ratios for the same lens.  If we consider the normalized errors from Eq.~(\ref{eq:prediction_errors}), accounting for both measurement and model uncertainties, we find a discrepancy $>3\sigma$ for all 9 mocks, and a discrepancy $>5\sigma$ for 5 out of 9 mocks between the EPL+shear prediction and the true flux ratios (see Figure~\ref{fig:summary_FR_prediction} for an illustration of these statistics).  Fold-like and cusp-like systems appear more susceptible to these false positives - this is expected since the quasar images are closer to the critical curve, and are therefore more sensitive to the presence of multipoles in the lens mass distribution.

To emphasize the importance of choosing the correct macromodel, we also fitted the 9 lensed systems with a macromodel that included free multipole parameters, i.e., EPL + shear + elliptical multipoles ($m=1$ + $m=3$ + $m=4$). For the example system from Figures~\ref{fig:with_mult_CAB_fit} and \ref{fig:with_mult_macromodel_fit}, we show the resulting flux-ratio posterior in Figure~\ref{fig:with_mult_FR_prediction}, which is now consistent with the true values, with slightly larger uncertainty compared to the EPL+shear model since we added some degrees of freedom. In fact, the false-positive flux-ratio anomaly detections all vanish when including multipoles in the macromodel: we show in Figure~\ref{fig:summary_FR_prediction} that all predictions are now accurate within $3\sigma$, with no discrepancy greater than $6\%$. This demonstrates that the most important consideration to avoid false-positive anomalies is to allow for enough flexibility in the lens model: as long as a macromodel is appropriately chosen, it can make unbiased flux-ratio predictions. 

In contrast, the alternative CAB method is more flexible and bypasses the need to specify the exact macromodel. The resampled estimates for the CAB-predicted flux ratios are consistent (within uncertainties) with the true flux ratios for all $9$ mock lenses - see Figure~\ref{fig:with_mult_FR_prediction} for an example, and Figure~\ref{fig:summary_FR_prediction} for a summary. This confirms that a CAB approach is robust to the presence of complexity in the lens mass distribution, in terms of flux-ratio anomaly detection. 
While the CAB formalism still relies on assumptions to construct the deflection field (here, that it behaves locally like an SIE+MST), they are much weaker than a macromodel (which imposes a common description for all 4 images), and they only relate to the extrapolation of local properties. The CAB always provides a valid description in the immediate vicinity of the lensed arc's center, and the matter is then how far from the center one can extrapolate using these assumptions, not if the assumptions are justified globally. In particular, CAB models do not require \textit{a priori} knowledge of what kinds of complexities are present in the deflector mass distribution, while perturbations allowed in a macromodel setup need to follow pre-specified patterns.

\subsection{Subhalo case: true-positive flux-ratio anomalies}
\label{subsubsec:with_DM_results}

Finally, we consider the 9 mock lenses for which the imaging data was generated with a DM subhalo creating a substantial flux-ratio anomaly for one of the quasar images. The presence of this individual subhalo does not perturb the lensed arc sufficiently to prevent the models from providing a good fit to the imaging data - we show an example of CAB fit in Figure~\ref{fig:with_DM_CAB_fit}.

Since the quasar images are not displaced by the subhalo significantly compared to the size of the ray-tracing region used to compute the magnification ratios  (see Section~\ref{subsubsec:mock_source}), we can directly compare the values with and without the subhalo ; in particular, we expect that an accurate smooth lens model will predict what the flux ratios would be without the subhalo. We find that both the macromodel (which matches the main deflector's underlying mass distribution in this case) and the CAB are able to make accurate predictions of the unperturbed flux ratios (see Figure~\ref{fig:with_DM_FR_prediction} for an example, and Figure~\ref{fig:summary_FR_prediction} for a summary over all $9$ mocks). 
This implies that CAB models do not absorb the local lensing perturbation from a DM subhalo, giving rise (if the flux perturbation is large enough to exceed the model-predicted and measurement uncertainties) to the same kind of flux-ratio anomalies from substructure that are widely discussed in the context of globally parametrized macromodels. We display an example of such a “true positive” flux-ratio anomaly in Figure~\ref{fig:with_DM_FR_prediction}: in this case, the perturbation of $\mathcal{F}_{23}$ by the subhalo is large enough that it would make the observed flux ratios statistically inconsistent with both the macromodel-predicted and CAB-predicted flux ratios. 

\newpage

\section{Summary \& Discussion}
\label{sec:discussion}

In this work, we have established Curved Arc Basis deflector models as a powerful alternative to globally parametrized macromodels in the context of flux-ratio prediction and anomaly detection in quadruply imaged quasars with extended arcs. 
Specifically:
\begin{itemize}
\vspace{-0.07cm}
    \item We have simulated mock lensed images with a variety of configurations and SNR values in the lensed arcs, and have compared the flux-ratio prediction capability of CAB models to a simple EPL+shear macromodel.
    \item We have established that CAB predictions accurately reproduce the expected flux-ratio values, within uncertainties that are typically $\sim3-5\%$ for our choice of cutouts.
    \item We have shown that, while a macromodel-based approach typically leads to smaller uncertainties in the model-predicted values since it imposes more physical assumptions on the mass profile, this lack of flexibility can lead to biased flux-ratio predictions if the macromodel is inadequate. In particular, when the simulated data has significant azimuthal complexity (in the form of multipole-like deviations from ellipticity) that is not captured by the macromodel, the latter can produce a false-positive anomaly in the flux ratios, whereas the proposed CAB implementation remains accurate.
    \item On the other hand, by injecting individual subhalos near quasar images, we have determined that CAB models do not fully absorb localized lensing perturbations, and are therefore capable of detecting flux-ratio anomalies that are actually produced by DM substructure, provided that they are large enough ($\gtrsim 10\%$) to exceed the model-predicted uncertainties.
\vspace{-0.07cm}
\end{itemize}
   
The importance of macromodel assumptions is already an acknowledged concern: current studies leveraging the flux-ratio anomaly statistics for DM substructure inference include several low-order multipoles in their macromodels for more flexibility \citep[e.g.][]{Gilman2022,Gilman2024, Keeley2024}, such that the relative likelihood of DM properties would already account for these types of perturbations. The true lens mass distribution, however, might deviate from a purely elliptical profile in ways that are not properly captured by the multipole expansion, either if the chosen parametrization is unsuited to describe physical expectations \citep[e.g., if the multipole formulation is assumed to have circular symmetry but the reference profile has non-negligible ellipticity,][]{Paugnat2025}, or if the mass profile displays other types of additional complexity, such as ellipticity gradients and twists in the isodensity contours \citep[e.g.,][]{Madejsky1990, VdV2022_twisting}, or a baryonic edge-on disk component \citep{Hsueh2016, Hsueh2017, Gilman2017}. The main benefit of the CAB modeling approach is therefore that it provides a robust alternative method to validate results obtained under the assumption of a macromodel. 

The CAB framework has its own drawbacks, mainly associated with its high flexibility. First, the statistical uncertainty on model-predicted flux ratios can be unreasonably high (compared to typical measurement uncertainties), especially if there is low SNR in the lensed arcs. We have circumvented this by resampling the noise maps and computing bootstrapped estimates, but this technique requires large computational resources. Second, CAB deflector models might find solutions that fit the imaging data well but are unphysical (e.g., with very different curvature centers between the independently modeled cutouts). To avoid this, our initial guesses for the CAB parameters were chosen close to the true values. In a real data modeling scenario, one would likely need to fit a decent macromodel first, then initialize the CAB parameters accordingly to allow for more flexibility. Numerical methods then have to keep a balance between exploring the parameter space extensively enough to allow the CAB model to adjust for complexity in the lens mass distribution, but selectively enough to stay in the realm of physical solutions. 

Ultimately, the CAB approach is intended to complement the traditional macromodel-based one: the idea is to bracket results between a constraining model with strong assumptions, and a less constraining model with enhanced adaptability. This can be expressed, essentially, as a choice between different priors on what the deflection field of a lensing galaxy looks like. The CAB would correspond to a broader, more conservative prior, potentially allowing unphysical configurations, while macromodels would represent narrower priors, based on our existing knowledge of galaxies but which risk overlooking some configurations. It is even possible to imagine an intermediate approach, by assuming that some CAB model parameters are correlated for a given lens, to enforce some physical assumptions while still allowing for more local flexibility. Testing different prior choices is important to ensure the robustness of DM constraints, in particular if a discovery of new physics is ever claimed from strong lensing.

We also remark that, while less rigid than in the macromodel case, the CAB models presented in this paper still rely on assumptions about the deflection field - namely, how the lensing properties (e.g., eigenvectors/eigenvalues of the lensing Jacobian) at the center of a lensed arc can be extrapolated in the vicinity of that point. In Appendix~\ref{App:systematic_source}, we establish that the $\sim 3\%$ flux-ratio systematic uncertainty estimated on high SNR mocks is likely due to the CAB assumptions adopted for Eq.~(\ref{eq:CAB_SIE+MST}). In theory, one could therefore improve this limiting source of uncertainty by devising a more sophisticated CAB deflector model, e.g. by allowing not only $\partial_{\rm tan}\lambda_{\rm tan} \neq 0$ but also higher-order derivatives of the eigenvalues to be non-zero, in particular along the tangential direction (see Appendix~\ref{App:systematic_source}). Given the current measurement uncertainties for the flux ratios \citep[$\sim2-10\%$, ][]{Nierenberg2024, Keeley2024}, the current model displays a reasonable performance, so we leave this to future work.

Finally, we emphasize that this paper is intended only as a proof of concept for the CAB approach in the context of flux-ratio anomalies. We have employed an ideal setup for our mock images, avoiding some of the difficulties that often arise when modeling real data. First, we have assumed a very simple PSF (Gaussian, with perfect knowledge of the parameters), whereas lensed quasar HST/JWST data display complex, extended PSF structure that needs to be accounted for, typically with iterative PSF reconstruction from a good initial model, otherwise there is a risk of biasing the lens modeling \citep[e.g.,][]{Shajib2022}. In particular, for JWST flux-ratio measurements, the uncertainties related to PSF modeling are still non-negligible \citep[e.g.,][]{Nierenberg2024}. This is expected to improve with the development of methods to generate better initial PSF estimates \citep[e.g.,][]{Williams2025}. In addition, we have performed our numerical experiment on mock systems with favorable scales, fixing the Einstein radius to $\theta_E=3^{\prime\prime}$ (in the upper part of the range for quad systems used in flux-ratio anomaly studies), in order to ensure that the quasar images were well-separated since this value is large compared to the PSF size and pixel scale. We have tested our method on a few examples with smaller values ($\theta_E=1.5^{\prime\prime}$), and we found that the CAB models still predict flux ratios accurately, albeit with larger statistical uncertainties. In our analysis, we have also neglected any light contribution from the main deflector galaxy. For real data, a lens light component would be added to the model - simple light profiles \citep[e.g., one or multiple elliptical Sérsics,][]{Schmidt2023, Nightingale2024} usually provide a reasonable subtraction of lens light at the location of the arcs. Therefore, this flux contribution might reduce the SNR in the lensed arcs, but we do not expect it to introduce significant systematic errors in the predicted flux ratios.
Finally, we have only considered individual subhalos to illustrate the case with true flux-ratio anomalies. More realistically, one would need to simulate entire populations for the line-of-sight field halos and for the subhalos of the main deflector, in particular to infer DM substructure properties with a forward-modeling approach \citep[e.g.,][]{Gilman2020, Gilman2024}. Integrating the CAB methodology in this framework will be the goal of a separate paper.


%% file: appendices.tex
\section{Signal-to-noise ratio and lensing information}
\label{App:SNR/I}

In our mock data, the noise maps are generated by randomly sampling background noise, plus photon shot noise described by Poisson statistics, such that, in a given pixel $i$ with a total measured count rate $s_i$, the SNR in that pixel is
\begin{equation}
    {\rm SNR}_{\rm pix}^{(i)} = \frac{s_i}{\sqrt{ \frac{s_i}{t_{\rm exp}}+\sigma_{\rm bkg}^2} }
\end{equation}
where $t_{\rm exp}$ is the exposure time \citep[here we fix $t_{\rm exp} = 1428\ {\rm s}$, to match a typical HST exposure in the F814W filter, e.g.,][]{Shajib2019}.
We can also define a per-pixel SNR for the lensed arcs only:
\begin{equation}
    {\rm SNR}_{\rm arc, pix}^{(i)} = \frac{s_{{\rm arc},i}}{\sqrt{ \frac{s_i}{t_{\rm exp}}+\sigma_{\rm bkg}^2} }
\end{equation}
where $s_{{\rm arc},i}$ is the flux contribution from the extended source (i.e., removing the quasar flux) in the pixel. A total SNR in the lensed arcs can then be calculated by summing the extended source contribution in all pixels with ${\rm SNR}_{\rm pix}>1$:
\begin{equation}
\begin{split}
    {\rm SNR}_{\rm arc} = \frac{S_{\rm arc}}{\sqrt{ \frac{S_{\rm tot}}{t_{\rm exp}}+ \sigma_{\rm bkg}^2N_{\rm arc}}} \text{ with } N_{\rm arc} &= \sum_{i} \mathbbm{1}_i\ , \\
    S_{\rm tot} = \sum_{i} s_i \mathbbm{1}_i \ , \text{ and } S_{\rm arc} &= \sum_{i} s_{{\rm arc},i} \mathbbm{1}_i .
\end{split}
\end{equation}
In this equation, $\mathbbm{1}_i=\mathbbm{1}( {\rm SNR}_{\rm arc, pix}^{(i)}>1)$, where $\mathbbm{1}$ is the indicator function, such that $N_{\rm arc}$ is the number of pixels where the lensed arc contribution is significant (compared to the background and quasar light), $S_{\rm tot}$ is the compounded signal in those pixels, and $S_{\rm arc}$ is the compounded contribution from the extended source only.

The SNR alone is not sufficient to quantify the information about the lensing field contained in the extended arcs as it should also depend on their spatial extent. In the most extreme case, a point source with extremely high SNR would not yield any constraints regarding the tangential or radial stretch. \cite{Tan2024} used a weighted SNR to account for this, but with a specific focus on the ability to constrain the power-law slope of the mass profile, while we are interested in the uncertainty on the magnifications at the quasar image locations. We adapt the idea of a weighting scheme and propose the following reasoning to quantify the lensing information in the context of flux-ratio prediction with a curved arc basis.

Let us consider a pixel with image plane coordinates $\vec{\theta}_i = \vec{\theta}_0 + d\vec{\theta}$  in the vicinity of the center of the arc $\vec{\theta}_0$ (see Figure~\ref{fig:CAB_illustration}). It maps back to the source plane coordinate
\begin{equation}
    \vec{\beta}_i \approx \vec{\beta}_0 + \mathbf{A}(\vec{\theta}_0)\cdot d\vec{\theta}
\end{equation}
such that we can write
\begin{equation}
    \begin{split}
       (\vec{\theta}_i-\vec{\theta}_0) \cdot \hat{e}_{\rm tan} &= \lambda_{\rm tan}^{-1}\  \beta_{i,\rm tan} \\
        (\vec{\theta}_i-\vec{\theta}_0) \cdot \hat{e}_{\rm rad}  &= \lambda_{\rm rad}^{-1}\  \beta_{i,\rm rad}
    \end{split}
\end{equation}
where $\beta_{i,k} \equiv (\vec{\beta}_i-\vec{\beta}_0) \cdot \hat{e}_k $ (with $k={\rm tan, rad}$) and $\hat{e}_{\rm tan}, \hat{e}_{\rm rad}$ are defined at the center of the arc. This point of the lensed arc can therefore be used to measure the stretch in the tangential ($\lambda_{\rm tan}$) or radial ($\lambda_{\rm rad}$) direction. Let us assume that the light distribution of the source is perfectly known, such that the intensity at pixel $i$ can be directly mapped to a position in the source plane with respect to the center of the light profile (i.e., $\beta_{i,\rm tan}$ and $\beta_{i,\rm rad}$ can be determined with negligible uncertainty). The ability to determine $\lambda_{\rm tan}$ and $\lambda_{\rm rad}$ is then limited by the precision $\delta_{\rm ref}$ to which we can measure distances in the image plane: typically, we can choose $\delta_{\rm ref} = \max (\delta_{\rm pix}, \sigma_{\rm PSF})$. Pixel number $i$ thus provides a constraint on $\lambda_k$ (with $k={\rm tan, rad}$) that has an uncertainty
\begin{equation}
    \delta\lambda_k = \frac{\delta_{\rm ref}}{|(\vec{\theta}_i-\vec{\theta}_0) \cdot \hat{e}_k |},
    \label{eq:delta_lamba_k}
\end{equation}
and the corresponding uncertainty on the log-magnification constraint at the center of the arc is 
\begin{equation}
    \delta\mathcal{M} = \frac{\delta \mu}{\mu} = \frac{\delta \lambda_{\rm tan}}{\lambda_{\rm tan}} + \frac{\delta \lambda_{\rm rad}}{\lambda_{\rm rad}}.
\end{equation}
This motivates the definition of the following quantity:
\begin{equation}
    \mathcal{P}_i \equiv \delta\mathcal{M}_i^{-1} =  \left[ \frac{\delta_{\rm ref}}{|r_i-s_{\rm tan}^{-1}|} +\frac{s_{\rm tan} \delta_{\rm ref}}{|\phi_i-\phi_0|} \right]^{-1}
    \label{eq:constraining_power}
\end{equation}
where $(r_i, \phi_i) \equiv \left( \| \vec{\theta}_i - \vec{\theta}_c\|, \ \arctan(x_i - \theta_c^x, y_i - \theta_c^y ) \right)$ are the coordinates of $\vec{\theta}_i$ and $(s_{\rm tan}^{-1}, \phi_0) $ are the coordinates of $\vec{\theta}_0$ in the polar coordinate system centered on the center of curvature $\vec{\theta}_c = \vec{\theta}_0 - s_{\rm tan}^{-1} \hat{e}_{\rm rad}$. We note that, while Eq.~(\ref{eq:delta_lamba_k}) was only valid in the vicinity of $\vec{\theta}_0$, we have extrapolated the definition of tangential/radial distance in the image plane by assuming that the curvature $s_{\rm tan}$ and the eigenvalues $\lambda_{\rm rad}, \lambda_{\rm tan}$ are constant along the arc, i.e., we performed the calculation under the simplest CAB deflector model described by Eq.~(\ref{eq:CAB_SIS+MST}), and assumed that this provides a good first-order approximation of the curved arc behavior even after generalizing to $\partial_{\rm tan}\lambda_{\rm tan}\neq0$. Based on the reasoning above, Eq.~(\ref{eq:constraining_power}) expresses the constraining power of pixel $i$, with respect to the prediction of the log-magnification at the center of the arc.

In reality, the source is not known, and the intensity measured at pixel $i$ has a noise term, so the $\beta_{i,k}$ would not be perfectly determined; but we can use this “ideal” constraining power to quantify the lensing information in a way that incorporates the spatial extent of the lensed arcs. For a cutout $\mathcal{C}_j$ defined around quasar image number $j$, we consider the SNR-weighted sum of the constraining power of each pixel in the cutout:
\begin{equation}
    \mathcal{P}_{\rm tot} (\mathcal{C}_j) = \sum_{i \in \mathcal{C}_j}  \mathcal{P}_i \times{\rm SNR}_{\rm arc, pix}^{(i)} \cdot \mathbbm{1}( {\rm SNR}_{\rm arc, pix}^{(i)}>1).
\end{equation}
In this work, we are interested in the flux ratios between all pairs of images, so the relevant metric is the lensing information in the more poorly constrained cutout, i.e.,
\begin{equation}
    \mathcal{P}_{\rm min} = \min_{1\leq j\leq4}\ \mathcal{P}_{\rm tot}( \mathcal{C}_j).
\end{equation}
Therefore, during the generation of mock data, for each case (baseline/with multipoles/with DM subhalo, see Section~\ref{subsec:mock_obs}) and each lens configuration (cross/cusp/fold), we simulated one lensed system with $\mathcal{P}_{\rm min} \approx 400$, one with with $\mathcal{P}_{\rm min} \approx 700$, and one with $\mathcal{P}_{\rm min} \approx 900$, in order to ensure diversity in the imaging data quality.

\section{Impact of cutout choice}
\label{App:cutout}

Choosing a cutout size is a matter of balancing three considerations: (a) having enough SNR in the cutout, (b) having pixels in the cutout with sufficient constraining power $\mathcal{P}_i$ (see Appendix~\ref{App:SNR/I}), such that $\lambda_{\rm rad}, \lambda_{\rm tan}$ can be constrained using distance measures that are not too limited by the PSF/pixel size, and (c) not including pixels that are too far from the center of the arc, such that the assumptions used to extrapolate the local lensing properties and form the CAB model (see Section~\ref{subsec:CAB_method}) do not introduce too much systematic error. 

\begin{figure*}
    \centering
    \includegraphics[width=0.97\textwidth]{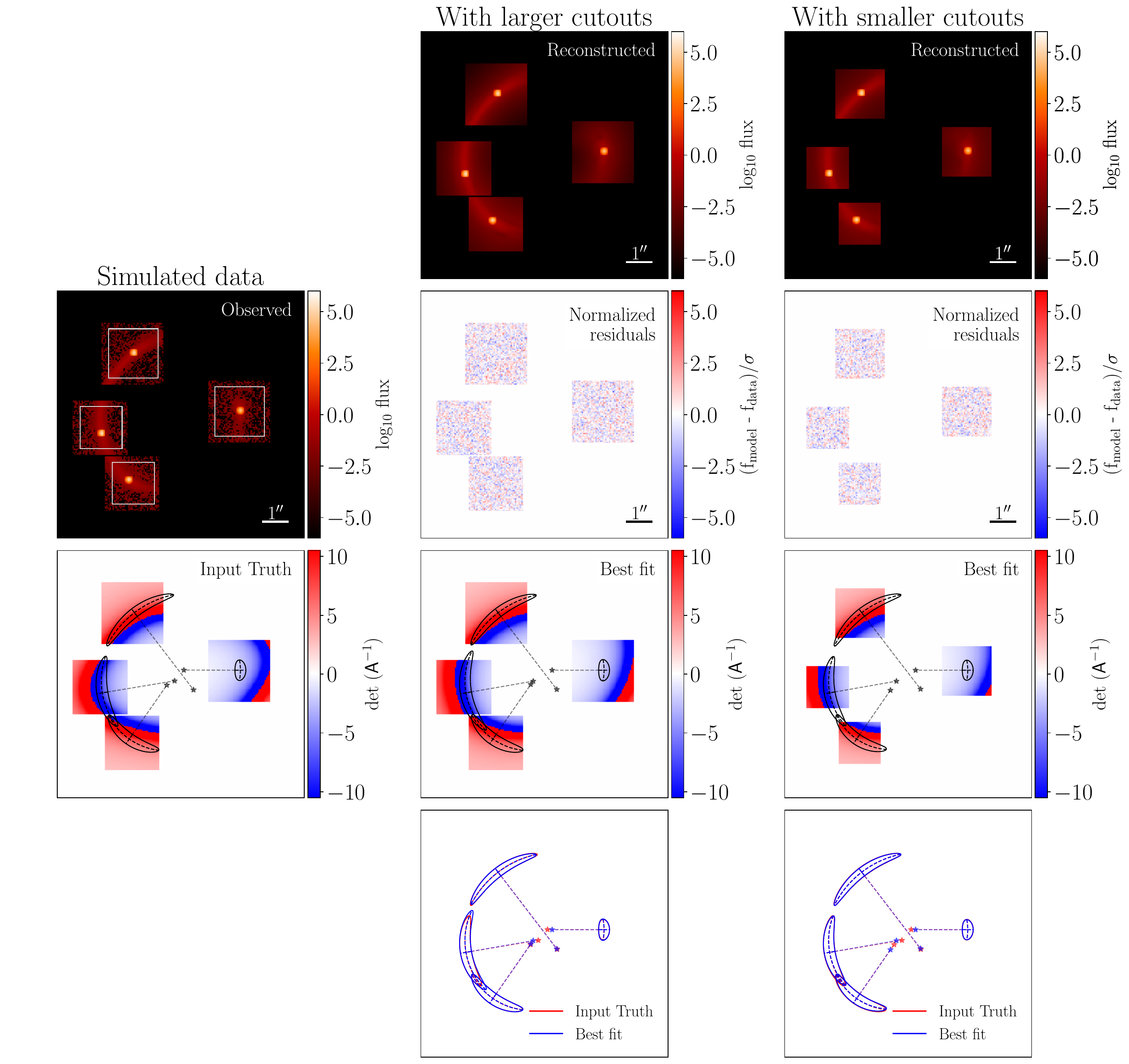}
    \caption{Example of CAB fit for a lensed system with multipoles, where the choice of cutout size has a notable impact on the flux-ratio prediction (this is a cusp-like system with ${\rm SNR}_{\rm arc}\approx 280$). Left column: Simulated imaging data for one realization (top), true magnification map and visual representation of the true CAB quantities at the quasar image locations (bottom). The plots are generated using the larger cutouts (with a default size of $50\times50$ pix, i.e. $2.5^{\prime\prime}\times2.5^{\prime\prime}$, subject to shrinking to avoid overlap - see Section~\ref{subsubsec:mock_data_quality}). The smaller cutouts (default size of $40\times40$ pix, i.e. $2.0^{\prime\prime}\times2.0^{\prime\prime}$) correspond to the white boxes in the imaging data. Middle column: For the larger cutout choice, reconstructed light for the best-fitting CAB model, normalized residuals in the imaging data, magnification map and CAB quantities at the quasar images for the best-fitting model, and comparison between best-fitting and true CAB quantities, respectively. Right column: Same as the middle column, for the smaller cutout choice.}
    \label{fig:impact_cutout_CAB_fit}
\end{figure*}

\begin{figure*}
    \centering
    \includegraphics[width=0.92\textwidth]{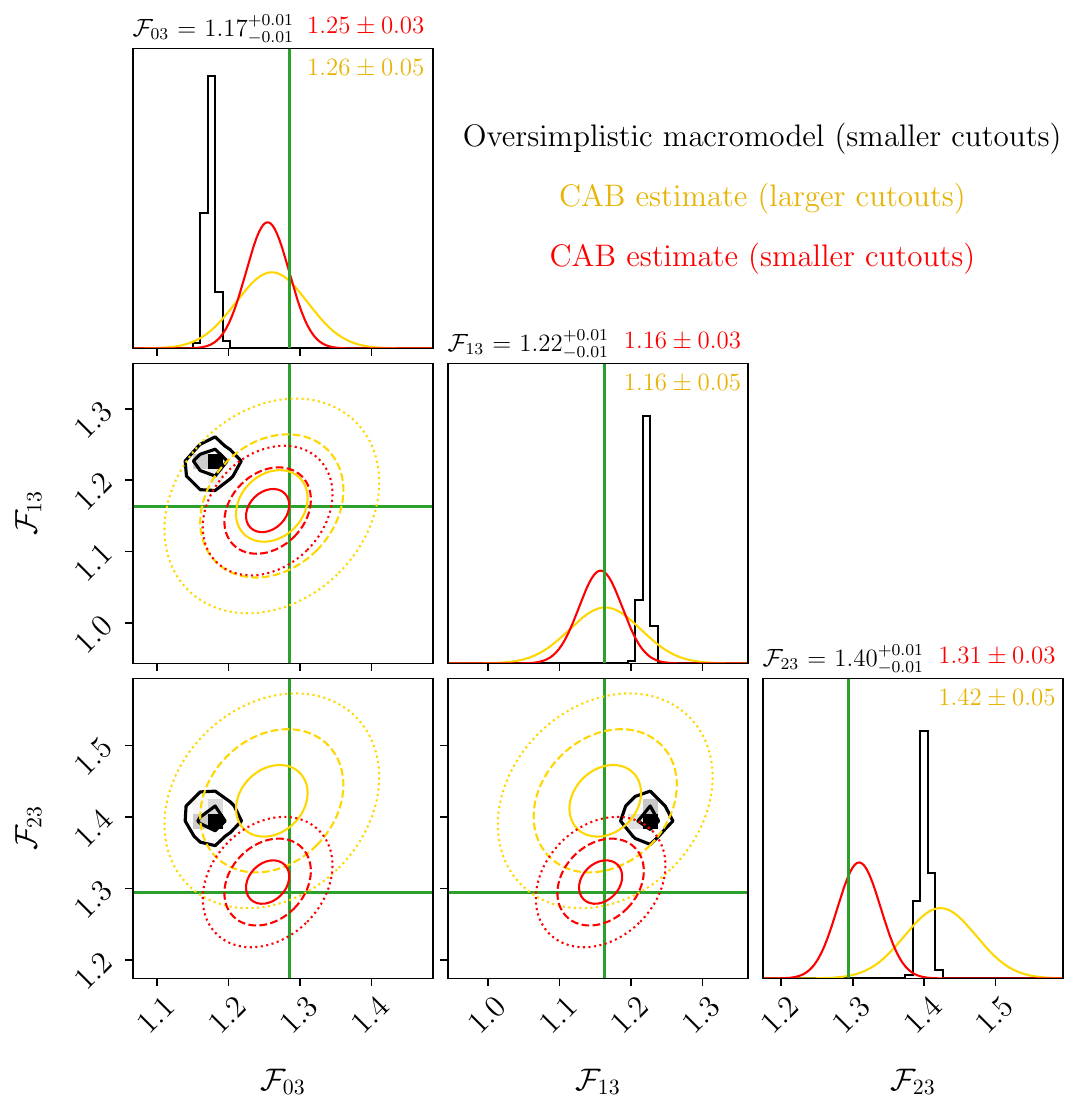}
    \caption{Same as Figures~\ref{fig:baseline_FR_prediction}, \ref{fig:with_mult_FR_prediction}, and \ref{fig:with_DM_FR_prediction}, but for the lensed system from Figure~\ref{fig:impact_cutout_CAB_fit}, which was fit for two different choices for the default cutout size. We note that the posterior distributions are broader for the larger cutouts, since the systematic uncertainty term estimated with high SNR mocks is larger (see Appendix~\ref{App:uncertainties}). The oversimplistic macromodel (EPL+shear) posteriors are consistent between the two cutout choices, so we only plot the one for the smaller cutouts.}
    \label{fig:impact_cutout_FR_prediction}
    \vspace{-0.1cm}
\end{figure*}

To illustrate the importance of point (c), we exhibit an example of lensed system in the complex case (i.e., with multipoles in the lens) where a change in the cutout size impacts the prediction of flux ratios by the CAB method described in Section~\ref{subsec:CAB_method}. This system was encountered during test runs and therefore does not belong to the sample of $9$ mocks discussed in Sections~\ref{subsubsec:with_mult_results}, but it was simulated in the same way. For the first run, the cutouts were given a default size of $50\times50$ pix (i.e., $2.5^{\prime\prime}\times2.5^{\prime\prime}$), before small adjustments (iteratively displacing and shrinking the cutouts) were made to avoid any overlap (see Section~\ref{subsubsec:mock_data_quality}). The methods described in Sections~\ref{subsubsec:macromodel_fit} and \ref{subsubsec:CAB_fit} were then applied to obtain a posterior distribution of log flux-ratios under an EPL+shear macromodel, and a CAB model (with resampling of the noise map), respectively. This modeling process was repeated with cutouts shrunk by $10$ pix (i.e., with a $40\times40$ pix or $2.0^{\prime\prime}\times2.0^{\prime\prime}$ default size), taking advantage of the high SNR. For both cutout sizes, we display an example of best-fitting CAB model for a noise map realization in Figure~\ref{fig:impact_cutout_CAB_fit}. The resulting CAB flux-ratio predictions are presented in Figure~\ref{fig:impact_cutout_FR_prediction}, along with the one for the EPL+shear macromodel. For the latter, the posterior is consistent between both cutout sizes, finding a $\gtrsim10\%$ and $\gtrsim5\sigma$ false-positive flux-ratio anomaly due to the azimuthal complexity (multipoles) not being captured by the macromodel (see Section~\ref{subsubsec:with_mult_results}). The CAB models perform better, with both predictions being consistent with the true flux ratios within $3\sigma$. For the larger cutout size, however, the uncertainties are predicted to be larger (in particular the expected systematic uncertainty is $\sim5\%$, see Table~\ref{tab:sys_uncert}), and yet the CAB-predicted value is still $\sim 3\sigma$ away from the truth, with a $\gtrsim10\%$ difference in one of the flux ratios. Shrinking the size of the cutouts solves this mild discrepancy, since the CAB predictions for the small cutouts are all within $1\sigma$ of the true values.

This experiment hints at the fact that the assumptions used to extrapolate local lensing properties into a complete CAB deflector model (see Section~\ref{subsec:CAB_method}) introduce some systematic errors, which risk biasing the flux-ratio prediction if the cutouts are too large, i.e., if they include pixels too far from the center of the arc. In fact, we show in Appendix~\ref{App:systematic_source} that the driving factor influencing the systematic uncertainty estimate (see Appendix~\ref{App:uncertainties}) is the angular size of the cutouts. This risk is increased for extreme cusp-like and fold-like configurations (where the arc is close to the critical curve), since the approximations underpinning CAB models break down more rapidly when angular deflections change significantly over short angular scales \citep[][]{Sengul2025}.

Therefore, when modeling real data, finding appropriate cutouts around each quasar image will possibly be non-trivial. A good first guess could be to adopt square cutouts with sides $\lesssim2\theta_E/3$, since this proved to be a good heuristic choice on simulated data (and we expect angular quantities to scale with the Einstein radius), but adjustments of the size and shape might be needed on a case-by-case basis to maximize their information content (i.e., minimize the statistical error) without introducing large systematic errors. A reasonable test would also be to vary the cutout size and check whether the CAB prediction is robust to this change. 


\section{Uncertainty budget}
\label{App:uncertainties}

We considered two sources of uncertainty on the log flux-ratio values predicted by the CAB deflector models:
\begin{itemize}
    \item the statistical error $\vec{\varepsilon}$ that comes from the noise in the image,
    \item the systematic error $\vec{S}$ that comes from the CAB method.
\end{itemize}
With this notation, the true log flux-ratios can be expressed as
\begin{equation}
    \mathcal{F}_{ij}^{\rm true} = \widehat{\mathcal{F}}_{ij} + S_{ij} + \varepsilon_{ij}
\end{equation}
where $\widehat{\mathcal{F}}_{ij}$ is the flux-ratio estimator (the median of log flux-ratios obtained from noise resampling, see Section~\ref{subsubsec:CAB_fit}). 
For convenience, we chose an arbitrary order to express the log flux-ratios between all pairs of images (and the corresponding errors) in vector form: 
\begin{equation}
    \vec{\mathcal{F}} = [\mathcal{F}_{01}, \mathcal{F}_{02}, \mathcal{F}_{03}, \mathcal{F}_{12}, \mathcal{F}_{13}, \mathcal{F}_{23}]^T
\end{equation}

For a single noise realization at low SNR, the error on flux ratios is completely dominated by the noise in the image. The posterior distribution on log flux-ratios, estimated using a Markov chain Monte Carlo (MCMC) algorithm, can be approximated as a multivariate Gaussian: $ \vec{\mathcal{F}}^{[1]} \sim \mathcal{N}(\vec{m}, \Sigma_{\rm MCMC})$, where $\Sigma_{\rm MCMC}$ is the covariance matrix. By resampling the noise in the image, we can compute a more robust estimator $\widehat{\mathcal{F}}$ (in our case, the median vector of the sample), and the variance of this estimator is expected to shrink as the number of samples $N_s$ increases. We assumed that the variance of this estimator behaved like the variance of the sample mean of data vectors independently sampled from a multivariate normal distribution, i.e.,
\begin{equation}
    \vec{\varepsilon} \sim \mathcal{N}\left( \vec{0}, \Sigma_{\rm stat}\right) \text{ with }  \Sigma_{\rm stat}=\frac{\Sigma_{\rm MCMC}}{N_s}
\label{eq:Sigma_stat}
\end{equation}
We warn that this approximation will break down if the dispersion of best-fitting values is too large (ie., if the posteriors from each noise realization are not consistent with each other). In that case, a more conservative estimate of the statistical uncertainty would be given by the empirical covariance from the sample of noise realizations: $\Sigma_{\rm stat}={\rm Cov}(\vec{\mathcal{F}},\vec{\mathcal{F}} )$. This situation can happen if the optimization routines struggle to find a global maximum in the likelihood. In our idealized setup, however, this did not happen since we used initial guesses that were close to the true values ; when modeling real data, a good strategy would likely be to initialize the CAB parameters with the values predicted by a good macromodel fit.

The systematic error $S_{ij}$ does not disappear, by definition, when resampling the noise map. It should average out, however, when considering many lens configurations, because the choice of image labeling is purely arbitrary: for every system where the CAB method overpredicts $\mathcal{F}_{ij}$, there should be a similar system where labels $i$ and $j$ are swapped, such that the CAB method underpredicts $\mathcal{F}_{ij}$ by the same amount. We thus modeled this systematic error term as a zero-mean multivariate Gaussian: $\vec{S} \sim \mathcal{N}( \vec{0}, \Sigma_{\rm sys} )$. 

Again, the variance of $S_{ij}$ should not depend on the choice of labels ($i,j$), so all the diagonal terms in $\Sigma_{\rm sys}$ should be equal. The systematic flux-ratio errors cannot be considered independent, however (i.e., $\Sigma_{\rm sys}$ is not diagonal): even if we change the labels, the systematic errors on $\mathcal{F}_{01}$ and $\mathcal{F}_{02}$ will always be positively correlated (because they both involve magnification $\mathcal{M}_{0}$), and the errors on $\mathcal{F}_{01}$ and $\mathcal{F}_{12}$ will always be anticorrelated (because they both involve magnification $\mathcal{M}_{1}$ with opposing signs). To account for this, we assumed the following form for the auto-covariance matrix:
\begin{equation}
    \Sigma_{\rm sys} = 
    \begin{bmatrix}
        \sigma_s^2 & c^2 & c^2 & -c^2 & -c^2 & 0\\
        c^2 & \sigma_s^2 & c^2 & c^2 & 0 & -c^2\\
        c^2 & c^2 & \sigma_s^2 & 0 & c^2 & c^2\\
        -c^2 & c^2 & 0 & \sigma_s^2& c^2 & -c^2\\
        -c^2 & 0 & c^2 & c^2 &\sigma_s^2 & c^2\\
        0 & -c^2 & c^2 & -c^2 & c^2& \sigma_s^2\\
    \end{bmatrix},
\end{equation}
where all the correlation (off-diagonal) coefficients are $\pm c$ or $0$ since, once again, the labels carry no physical meaning (the only fact that matters is whether there is a repeated label in a pair of flux ratios).

To determine the coefficients $\sigma_s$ and $c$, we simulated ${N_{\rm mock}\sim 100}$ mock lenses with $\mathcal{P}_{\rm min} \sim 5 \times 10^5$ (see Appendix~\ref{App:SNR/I}), which resulted in extremely high SNR in the arcs (typically, ${\rm SNR}_{\rm arc} \gtrsim 10^3$), such that $\varepsilon_{ij}$ became negligible. In order to avoid systematics due to imperfect source modeling with our fixed number of shapelets (see Section~\ref{subsubsec:source_model}), we generated the mock imaging data using an elliptical Sérsic fit to the COSMOS galaxy that had been drawn, instead of the exact surface photometry. After fitting the imaging data, the difference between the true and predicted flux ratios $S_{ij}= \mathcal{F}_{ij}^{\rm true} - \widehat{\mathcal{F}}_{ij}$ for each lens was used to compute the following estimates: 
\begin{equation}
\begin{split}
    \sigma_s ^2 &= \frac{1}{6N_{\rm mock}} \sum_{p=1}^{N_{\rm mock}} \sum_{i<j}(S_{ij}^{(p)})^2\\
    c^2 &= \frac{1}{24N_{\rm mock}} \sum_{p=1}^{N_{\rm mock}} \sum_{i<j} \sum_{k<l} \eta_{ijkl} \cdot S_{ij}^{(p)}S_{kl}^{(p)},
\end{split}
\end{equation}
    where $\eta_{ijkl} = \delta_{ik}+\delta_{jl}-\delta_{il}- \delta_{jk}-2\delta_{ik}\delta_{jl}$ selects pairs of flux ratios with a single repeating label and gives them the appropriate sign, and $\delta_{ij}$ is the Kronecker delta symbol.

We used this technique to estimate the level of systematic uncertainty that could be expected from the CAB method in each mock lens case (baseline, with multipoles, with subhalo). For the case with multipoles, instead of imposing a significant boxyness/diskyness (with $a_4=\pm1.5\%$ and $\varphi_4=0$) like in Section~\ref{subsubsec:mock_lens}, we adopted a sampling distribution informed by physical expectations \citep[$a_4 \sim \mathcal{N}(0, 0.01)$ and $\varphi_4 \sim \mathcal{U}(-\pi/8, \pi/8)$, e.g., ][]{Hao2006, Gilman2024, Paugnat2025}, to obtain an uncertainty estimate more representative of what one can expect in an actual lensed quasar sample. We report the results in Table~\ref{tab:sys_uncert}, presenting the systematic uncertainty on individual log flux-ratios ($\sigma_s$), as well as the Pearson Correlation Coefficient (PCC) between log flux-ratios with a label in common ($c^2/\sigma_s^2$). We find consistent values between all three cases, with $c^2/\sigma_s^2 \sim 0.25$ and $\sigma_s\sim 0.03$. We further investigate the source of this $\sim 3\%$ systematic uncertainty in Appendix~\ref{App:systematic_source}, concluding that it is likely due to the assumptions used to build the CAB deflector model, and that this number is mostly impacted by the physical size of the cutouts used when fitting the imaging data.

\addtolength{\tabcolsep}{10pt}
\begin{table*}
\caption{Systematic uncertainty $\sigma_s$ on individual log flux-ratios under a CAB model, and Pearson Correlation Coefficient between log flux-ratios with a common label, for a variety of mock lens cases, CAB assumptions, and data characteristics. These values are estimated using simulated quads with high SNR, such that statistical uncertainty can be neglected.}
\label{tab:sys_uncert}
\resizebox{\textwidth}{!}{
\begin{tabular}{cccccc}
\hline
\textrm{Default cutout}& \textrm{PSF width}& \textrm{Pixel size} &  & Systematic & PCC \\
\textrm{size ($^{\prime\prime}$)}& \textrm{$\sigma_{\rm PSF}$ (mas)}& \textrm{$\delta_{\rm pix}$ (mas)} & Case & uncertainty $\sigma_s$ & ($= c^2/\sigma_s^2$) \\
  \hline  \hline
2.0 & 80 & 50 & \text{Baseline} & 0.030 &  0.24 \\
2.0 & 80 & 50 & \text{With multipoles$^*$ (\textdagger)} & 0.031 &  0.25 \\
2.0 & 80 & 50 & \text{With DM subhalo} & 0.034 &  0.21 \\
\hline
2.5 & 80 & 50 & \text{With multipoles$^*$} & 0.050  &  0.22 \\
2.5 & 120 & 50 & \text{With multipoles$^*$} & 0.045 &  0.23 \\
2.5 & 80 & 25 & \text{With multipoles$^*$} & 0.045 &  0.21 \\
2.5 & 80 & 50 & \text{With multipoles$^*$, $\partial_{\rm tan}\lambda_{\rm tan}=0$} & 0.061 &  0.21 \\
\hline
\end{tabular}}\\
\footnotesize{$ $\\$^*$The $m=4$ multipole values are sampled with $a_4 \sim \mathcal{N}(0, 0.01)$ and $\varphi_4 \sim \mathcal{U}(-\pi/8, \pi/8)$, instead of being fixed like in Section~\ref{subsubsec:mock_lens}.}
\end{table*}
\addtolength{\tabcolsep}{-11pt}

For the final uncertainties, assuming Eq.~(\ref{eq:Sigma_stat}) for the statistical part, we can write $\vec{\mathcal{F}}^{\rm true} \sim \mathcal{N} \left( \widehat{\mathcal{F}}, \frac{\Sigma_{\rm MCMC}}{N_s} + \Sigma_{\rm sys} \right)$. In particular, the marginalized distribution on a single log flux-ratio is a Gaussian, with standard deviation $\sigma_{ij}$ given by:
\begin{equation}
    \sigma_{ij}^2 = \frac{(\sigma_{ij}^{[1]})^2}{N_s} + \sigma_s^2,
\end{equation}
where $\sigma_{ij}^{[1]}$ is the uncertainty for a single noise realization, determined with the MCMC exploration of the posterior. Typically, the final uncertainties for the log flux-ratios are of order $\sigma_{ij} \sim 0.03-0.05$, i.e.,  $3-5\%$ uncertainty on the flux ratios, comparable to the uncertainties of the warm dust flux-ratio measurements \citep{Nierenberg2024, Keeley2024}.

We note that this uncertainty budget does not include PSF uncertainties (since we assumed a Gaussian PSF with fixed width) nor uncertainty due to source light reconstruction (in our case, how well the surface photometry of the quasar host galaxy can be captured by an elliptical Sérsic plus a shapelet basis expansion). For simplicity, and to avoid overfitting, we chose the smallest number of shapelets ($n_{\rm max}=5$) that can generically fit the data down to the noise level. In practice, the number of shapelets can be determined more rigorously by choosing, for each lens, the $n_{\rm max}$ value that minimizes the Bayesian information criterion (BIC), ensuring an optimal level of complexity in the source reconstruction \citep[e.g.,][]{Gilman2024}. Consequently, we expect this to be a negligible source of error for the flux ratios, which is why we used simplified sources (with elliptical Sérsic profiles) to estimate the systematic uncertainty. The PSF, on the other hand, is expected to be a significant source of uncertainty for lens modeling on real data \citep[e.g.,][]{Nierenberg2024, Williams2025}, often requiring sophisticated reconstruction techniques. We have purposefully ignored such complications in this analysis, and leave the estimation of flux-ratio uncertainties due to the PSF to future work.  

\section{Investigation on the source of systematic error}
\label{App:systematic_source}

In order to determine the factor(s) driving systematic errors in the CAB flux-ratio prediction, we re-estimated the systematic uncertainty level $\sigma_s$ and the PCC (employing the same technique with high SNR mocks as in Appendix~\ref{App:uncertainties}) after changing some simulation parameters: the physical size of the cutouts, the PSF size, and the pixel size. The results are reported in Table~\ref{tab:sys_uncert}. We find that the distribution of systematic errors does not depend (or only depends mildly) on the PSF and pixel size, but changes significantly with the angular size of the cutouts (with a $\sim60\%$ increase in $\sigma_s$ when making the default  cutouts bigger by $0.5^{\prime\prime})$. This matches the observation made in Appendix~\ref{App:cutout}, where we found that excessively large cutouts could lead to biases in the CAB flux-ratio prediction, an issue that could be resolved by shrinking the size of the fitting region.

We therefore suspect that extrapolation assumptions underpinning the CAB models are responsible, since these would be more involved as pixels farther away from the center of the arc are included. To confirm this, we re-estimated $\sigma_s$ after fixing $\partial_{\rm tan}\lambda_{\rm tan}=0$, that is, we added an assumption to the CAB model, employing the model from Eq.~(\ref{eq:CAB_SIS+MST}), which is a special case of the more general one from Eq.~(\ref{eq:CAB_SIE+MST}). We find that $\sigma_s$ subsequently increases by $\sim20\%$ (see Table~\ref{tab:sys_uncert}), confirming that extrapolation assumptions are the driving source of systematic error in CAB flux-ratio prediction.

\begin{figure}
\centering
    \includegraphics[width=0.95\columnwidth]{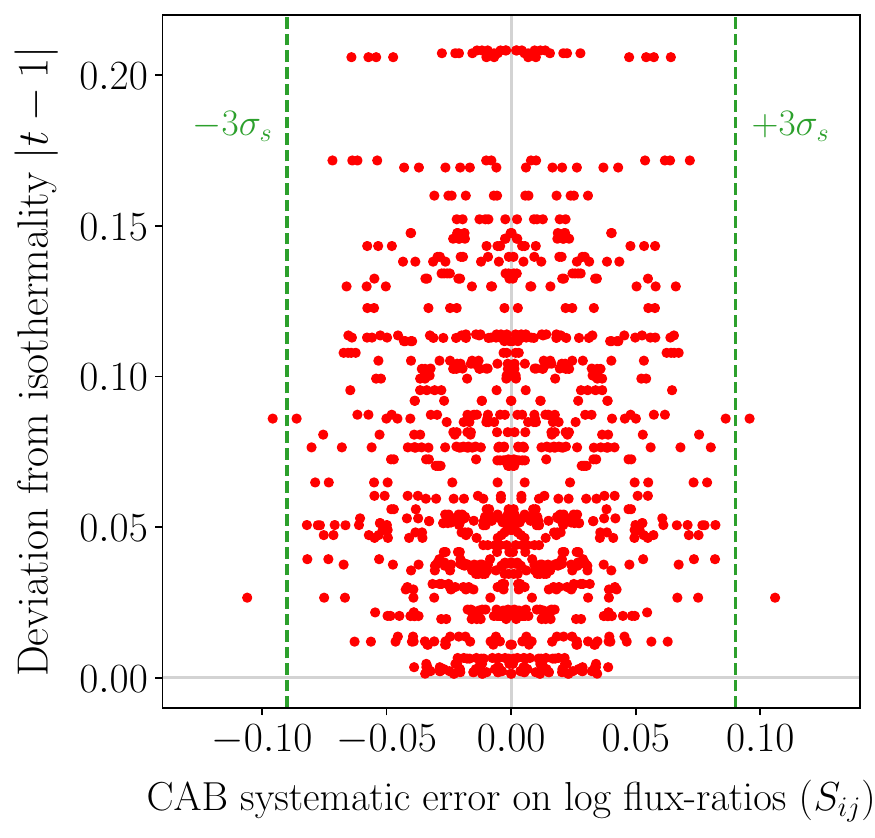}
    \caption{Systematic errors $S_{ij}$ on log flux-ratios due to applying CAB models to high SNR mock quads, plotted against the deviation from isothermality $|t-1|$ in the EPL profile of the mock lens. The mocks used here belong to the run labeled (\textdagger) in Table~\ref{tab:sys_uncert}, with the corresponding $\pm3\sigma$ limits shown as green dashed lines. }
    \label{fig:slope_vs_pred_error}
\end{figure}

By analogy with the generalization from Eq.~(\ref{eq:CAB_SIS+MST}) to Eq.~(\ref{eq:CAB_SIE+MST}) \citep{Birrer_CAB}, 
a natural idea to improve CAB models is to include a first-order differential for the radial stretch, i.e., to consider CAB models with  $\partial_{\rm rad}\lambda_{\rm rad}\neq0$. This could probably be constructed using a local EPL+MST deflection field (since Eq.~(\ref{eq:CAB_SIE+MST}) is equivalent to a local SIE+MST, and the EPL is a generalization of the SIE with free radial slope $t$). If the radial approximation $\partial_{\rm rad}\lambda_{\rm rad}=0$ was responsible for systematic errors in the CAB flux-ratio prediction, however, we would expect larger errors in EPL lenses when the radial slopes are far from isothermal - that is, the distribution of $S_{ij}$ should become broader as the deviation from isothermality $|t-1|$ increases. This effect is not observed in our high SNR mock data, as illustrated in Figure~\ref{fig:slope_vs_pred_error}  for one of the sets of simulation parameters from Table~\ref{tab:sys_uncert}. We conclude that greater radial flexibility is unlikely to improve the performance of CAB flux-ratio prediction, implying that the systematic uncertainty is driven by the extrapolation assumptions in the tangential direction.

We note that the findings of this Appendix imply that the value $\sigma_s\sim 0.03$ reported in Section~\ref{subsubsec:CAB_fit} and Appendix~\ref{App:uncertainties} is specific to our choice of default cutout size ($2^{\prime\prime}\times2^{\prime\prime}$). If, like suggested in Appendix~\ref{App:cutout}, the cutouts are tailored to each real data case, the high SNR simulation would need to be rerun to obtain a systematic uncertainty value that matches these adjustments. This might become convoluted if the cutouts have complex shapes, or very different sizes.
Still, we have shown that, with a very straightforward approach, a systematic uncertainty of $\sim 3\%$ and final uncertainties of $\sim3-5\%$ on flux-ratio prediction with CAB models is an achievable objective.